\documentclass[aps,pre,reprint,superscriptaddress,twocolumn]{revtex4-1}

\usepackage{graphicx}
\usepackage{amsfonts}
\usepackage{amsmath}
\usepackage{amssymb}
\usepackage{natbib}
\newcommand {\e}[1]{\mathrm{~#1}}    
\newcommand {\E}[1]{\cdot 10^{#1}}		

\graphicspath{{FigureWork/}}

\begin{document}

\title{Nonlinear decoding of a complex movie from the mammalian retina}

\author{Vicente Botella-Soler}
\affiliation{Institute of Science and Technology Austria, Klosterneuburg, Austria}
\author{St\'{e}phane Deny}
\author{Olivier Marre}
\affiliation{Institut de la Vision, INSERM UMRS 968, UPMC UM 80, CNRS UMR7210, Paris, France}

\author{Ga\v{s}per Tka\v{c}ik}
\affiliation{Institute of Science and Technology Austria, Klosterneuburg, Austria}

\begin{abstract}
Retinal circuitry transforms spatiotemporal patterns of light into spiking activity of ganglion cells, which provide the sole visual input to the brain. Recent advances have led to a detailed characterization of  retinal activity and stimulus encoding by large neural populations. The inverse problem of decoding, where the stimulus is reconstructed from spikes, has received less attention, in particular for complex input movies that should be reconstructed ``pixel-by-pixel''. We recorded around a hundred neurons from a dense patch in a rat retina and decoded movies of multiple small discs executing mutually-avoiding random motions. We constructed nonlinear (kernelized) decoders that improved significantly over linear decoding results, mostly due to their ability to reliably separate between neural responses driven by locally fluctuating light signals, and responses at locally constant light driven by spontaneous or network activity. This improvement crucially depended on the precise, non-Poisson temporal structure of individual spike trains, which originated in the spike-history dependence of neural responses.
Our results suggest a general paradigm in which downstream neural circuitry could discriminate between spontaneous and stimulus-driven activity on the basis of higher-order statistical structure intrinsic to the incoming spike trains.
\end{abstract}


\maketitle

Decoding plays a central role in our efforts to understand the neural code~\cite{spikesbook,oram+al_1998,georgopoulos+al_1986,kay+al_2008}. While statistical analyses of neural responses can be used to directly estimate~\cite{strong+al_1998,archer+al_2013} or bound~\cite{tkacik+al_2014} the information content of spike trains, such analyses remain agnostic about what the encoded bits might mean or how they could be read out~\cite{borst+theunissen_1999}. In contrast, decoding provides an explicit computational procedure for recovering the stimulus from recorded single-trial neural responses, allowing us to ask not only ``how much'', but also ``what''  the neural system encodes~\cite{quiroga+panzeribook}. This is particularly relevant when a rich stimulus is represented by a large neural population---a regime which is increasingly accessible due to recent experimental progress, and the regime that we explore here.

One of the hallmarks of large-scale neural activity is the presence of spontaneous and persistent spiking~\cite{Destexhe+Contreras_2006,major+tank_04}. While commonly discussed in a cortical context~(e.g,~\cite{Ringach_2009,tsodyks+al_99}), similar activity can also be observed in the sensory periphery~\cite{kuffler+al_1957,troy+lee_94,shlens+al_06,freeman+al_08}. From the viewpoint of any downstream information processing, either by the brain itself or by the decoding algorithms we construct, spontaneous activity presents a confound: if erroneously interpreted as having been caused externally, the system will ``hallucinate'' nonexistent stimuli and will likely respond inappropriately. Importantly, similar confounds can also happen locally when stimuli are high dimensional, e.g., in parts of a visual scene where light intensity does not fluctuate.  Can the neural activity be disambiguated so as to enable reliable stimulus representations, and ultimately percepts, even in presence of spontaneous firing? More generally, how can complex stimuli be reconstructed from neural activity?  We address these questions by decoding the outputs of a mammalian retina.

Decoding from large populations presents a significant technical challenge due to its intrinsic high dimensionality. Past work has predominantly addressed this problem using two approaches. In the first approach, one only presents stimuli that have simple, low-dimensional representations, in order to turn decoding into a tractable fitting (e.g., angular velocity of a moving pattern~\cite{bialek+al_1991}, luminance flicker~\cite{warland+al_1997}, 1D bar position~\cite{marre+al_2015}, etc.) or classification problem (e.g., shape identity~\cite{schwartz+al_2012}, a small set of orientations or velocities~\cite{frechette+al_2005}, etc.). It is unclear, however, how results for simple stimuli can be generalized to naturalistic stimuli even in principle, as the latter have no low-dimensional representation and, furthermore, the retinal responses are nonlinear. In the second approach, one first builds a probabilistic encoding model, followed subsequently by model-based inference of the most likely stimulus given the observed neural responses~\cite{pillow+al_2008,meytlis+al_2012,nichols+al_2013}. Theoretically, this procedure is possible for any stimulus, but in practice model inference is feasible only if it incorporates strong dimensionality reduction assumptions (e.g., that neurons respond to a linear projection of the stimulus).
Here we demonstrate a third alternative, where a complex and dynamical stimulus is reconstructed from the output of the mammalian retina directly, by means of large-scale kernelized regression~\cite{paiva+al_2009}. Retina is an ideal experimental system for such a study, because it permits stable recordings from large, diverse, local populations of neurons under controlled stimulation, where even simultaneous neural spiking events can be sorted reliably~\cite{mare+al_2012}.

We start by performing linear decoding from the entire recorded retinal ganglion cell population, to separately reconstruct the temporal light intensity trace at each spatial location in the stimulus movie. When using sparse regularization, we extract and subsequently analyze  ``decoding fields,'' the decoding counterpart of the cells' receptive fields. We next examine nonlinear decoding using kernel ridge regression (KRR), which provides a substantial increase in performance over linear decoding, and isolate spike train statistics that the nonlinear decoder is making use of. We conclude by examining how these statistics arise in generative models of spike trains and suggest that they might be essential for separating stimulus-driven from spontaneous activity.

\section{Results}

\subsection{Sparse linear decoding of a complex movie}
\begin{figure*}
\centerline{\includegraphics[width=\textwidth]{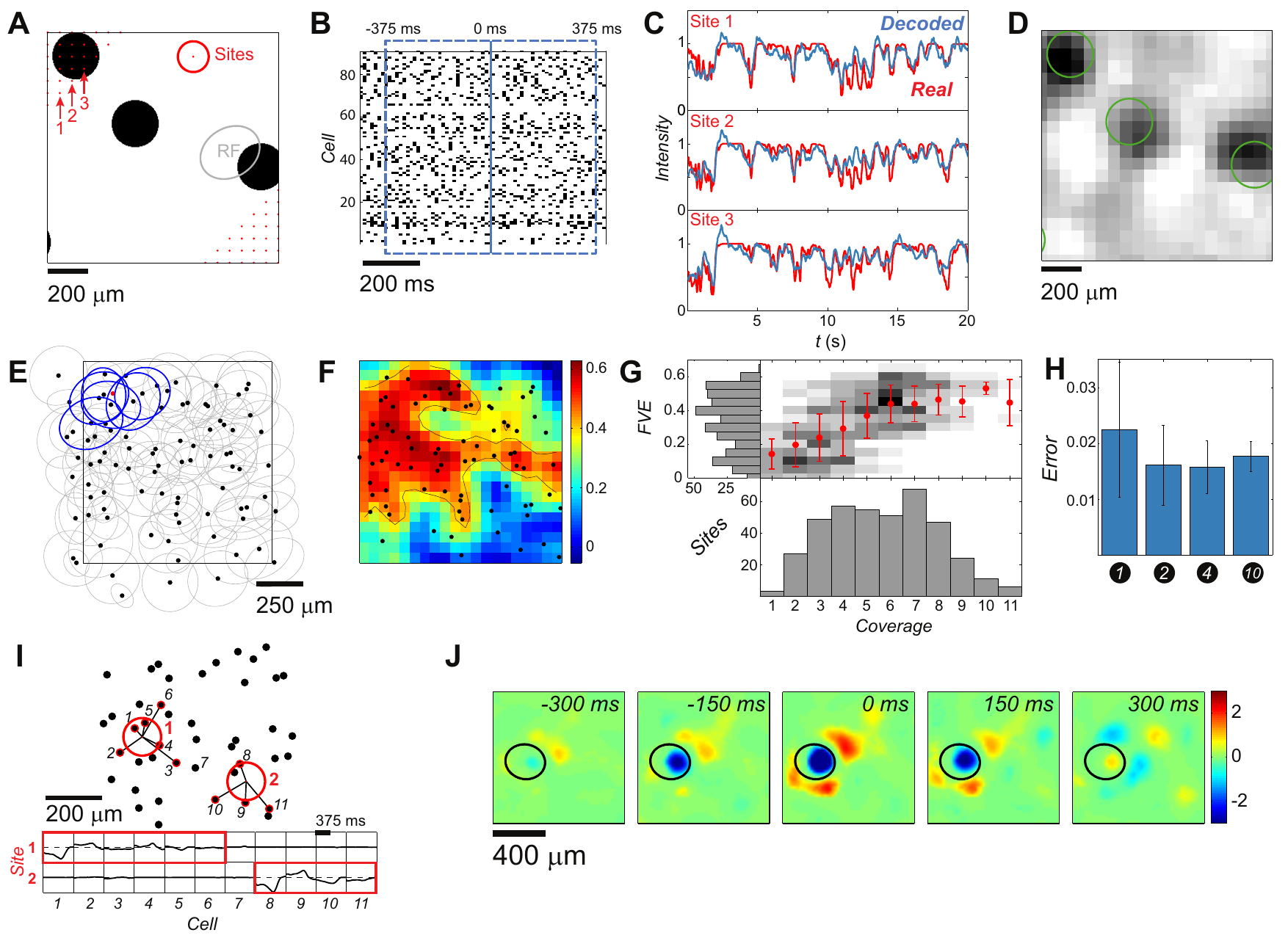}}
\caption{\textbf{Linear decoding of a complex movie.}
\textbf{A}: An example stimulus frame. At each site (red dots = partially shown 20$\times$20 grid) the stimulus was convolved with a spatial gaussian filter (red circle = $1\sigma$). Typical RGC receptive field center size shown in gray.
\textbf{B}: Responses of 91 RGCs with $750\e{ms}$ decoding window overlaid in blue.
\textbf{C}: Three example luminance traces (red) and the linear decoders' predictions (blue). 
\textbf{D}: Decoded frame (same as in \textbf{A}) reconstructed from 20$\times$20 separately decoded traces. Disc contours of the original frame shown for reference in green.
\textbf{E}: RF centers of the 91 cells (black dots = centers of fitted ellipses). RF centers overlapping a chosen site (red dot) are highlighted in blue.
\textbf{F}: Performance of the linear decoders across space, as Fraction of Variance Explained (FVE). Black dots as in \textbf{E}; black contour is the boundary $FVE=0.4$. 
\textbf{G}: Performance of the linear decoders (FVE) across sites as a function of cell coverage (grayscale = conditional histograms, red dots = means, error bars = $\pm$ SD).
\textbf{H}: Average decoding error across sites (MSE $\pm$ SD) of 10-disc-trained decoders, tested on withheld stimuli with different numbers of discs.
\textbf{I}: Cells (black dots = RF center positions) contributing to the decoding at two example sites (red circles); decoding filters shown below. For each site, contributing cells (highlighted in red and joined to the site) account for at least half of the total L1 norm.
\textbf{J}: Decoding field of a single cell (here, evaluated over a denser 50$\times$50 grid and normalized to unit maximal variance); the cell's RF center shown in black.}\label{Figure1}
\end{figure*}

We recorded the spiking activity of $N=91$ ganglion cells from a 1 mm$^2$ patch of the rat retina, while presenting a complex and dynamical stimulus that consisted of $1, 2, 4$ or $10$ randomly moving black discs on a bright background (Fig.\,\ref{Figure1}A and Methods). Our goal was to reconstruct the light intensity as a function of time (``luminance trace'') at a grid of $20\times 20$ spatial positions (``sites'') uniformly tiling the stimulus frame. Stimulus features (here, disc size) were smaller than the receptive field center of a typical recorded RGC, making the decoding task non-trivial. 

To estimate the luminance trace at any given time, we trained a separate sparse linear decoder for each site on a $750\e{ms}$ sliding window of the complete spiking raster, shown in Fig.\,\ref{Figure1}B, and represented as spike counts in $\Delta t = 12.5\e{ms}$ time bins (see Methods). While each decoder in principle had access to all neural responses, our sparse (L1) penalty on decoding weights ensured that the majority of the weights corresponding to redundant or non-informative neural responses for each site were zeroed out, yielding interpretable results which we describe in detail below. When trained on the 10-disc stimulus, this procedure predicted well the luminance traces across individual sites on withheld sections of the stimulus (Fig.\,\ref{Figure1}C), allowing us to reconstruct the complete movie (Fig.\,\ref{Figure1}D).

We expected the performance of our decoder to depend strongly on  local coverage, i.e., on the number of recorded cells whose receptive field centers overlap a given site. Coverage amounted to about six cells on average and exhibited substantial spatial heterogeneity, as shown in Fig.\,\ref{Figure1}E. The quality of our movie reconstruction, measured locally by  ``fraction of variance explained'' (FVE, see Methods), showed similar spatial variation (Fig.\,\ref{Figure1}F) which correlated strongly with coverage (Fig.\,\ref{Figure1}G), and saturated at $\geq 6$ cells. In what follows, we restrict our analyses to sites with good coverage that pass a threshold of $\mathrm{FVE}\geq 0.4$. Despite the high dimensionality of this regression problem (decoders have $\sim 5\E{3}$ parameters per site), sparse regularization ensured uniformly good performance even when  tested on out-of-sample stimuli with varying number of discs (Fig.\,\ref{Figure1}H).

To analyze how rich stimuli are represented by a population of ganglion cells with densely overlapping receptive fields, we examined the resulting decoding weights in detail. We found that stimulus readout was surprisingly local. As illustrated for two example sites in Fig.\,\ref{Figure1}I, only a few cells whose receptive field centers were in close proximity to the respective sites were assigned non-negligible decoding weights. This was true in general: on average $5.4\pm 2.8$ cells, whose RF centers were all located within $200\e{\mu m}$ of the decoded site, contributed to the luminance trace reconstruction; cells beyond this spatial scale contained no decodable information (SI~Fig.\,1, 2). 

Our framework also allowed us to construct a ``decoding field'' for every cell (Fig.\,\ref{Figure1}J). A decoding field represents an impulse response of the decoder, i.e., an additive  contribution to the stimulus reconstruction for every spike emitted by a particular cell. Because neural encoding is strongly nonlinear, there is no \emph{a priori} reason for the similarity between receptive and decoding fields, especially when, as here, they were inferred under different stimulus conditions. We nevertheless found a remarkably good correspondence: spatial locations and sizes of the decoding and receptive fields coincided for all cells (SI~Fig.\,3), with decoding fields furthermore exhibiting a clear center-surround-like structure. Taken together, these and supplementary results (SI~Figs.\,4, 5, 6) suggest that retinal responses to complex stimuli can be read out in a highly stereotyped, structured, and local manner.

\begin{figure}
\centerline{\includegraphics{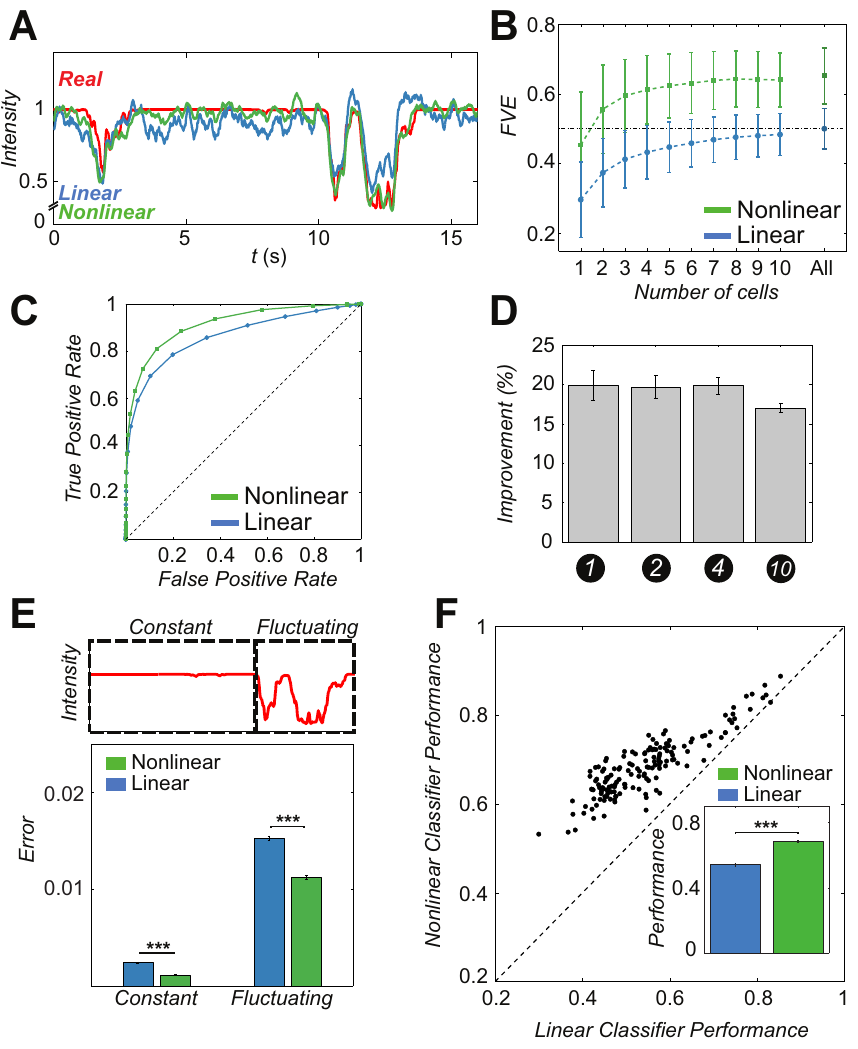}}
\caption{\textbf{Nonlinear decoding outperforms linear decoding.}
\textbf{A}: Luminance trace (red) with linear (blue) and nonlinear (green) predictions.
\textbf{B}: Average decoder performance ($\pm$ SD across sites), achievable using increasing numbers of cells with highest L1 filter norm. For nonlinear decoding, ``All''  is the optimal subset that maximizes performance (SI~Fig.\,7).
\textbf{C}: Average ROC across all testing movie frames.
\textbf{D}: Fractional improvement (average $\pm$ SEM across sites) of nonlinear versus linear decoders for test stimuli with different numbers of discs. All decoders were trained only on the 10-disc stimulus.
\textbf{E}: Decoding error (MSE; average $\pm$ SEM across sites) in fluctuating and constant epochs is significantly larger for linear decoders (p$<$0.001). 
\textbf{F}: Performance (F-score) of linear and nonlinear classifiers for different sites (black dots). Inset: average ($\pm$ SEM) over sites is significantly different (p$<$0.001). 
 }\label{Figure2}
\end{figure}

\subsection{Nonlinear decoding outperforms linear decoding}

Could nonlinear decoding improve on these results? A tractable method that extends linear regression into the nonlinear domain is kernel ridge regression (KRR), which we applied to our recordings using Gaussian kernels of cross-validated width (see Methods)~\cite{Park2013}. Importantly, the success of the nonlinear decoder crucially depended on the proper selection of local groups of cells relevant for each site, as identified by linear decoding: its sparse (L1) regularization acted as ``feature selection'' for the nonlinear problem (Methods, SI~Fig.\,7). Nonlinear decoder could then make use of higher-order statistical dependencies within and between the selected spike trains to achieve high performance.

Figure~\ref{Figure2}A shows a luminance trace at one of the example sites, together with its linear and nonlinear reconstruction. Nonlinear decoder tracks better the detailed structure of luminance troughs, which occur when discs cross the site, as well as exhibiting smaller fluctuations when no discs are crossing the site and the true luminance trace is therefore constant. This is reflected in a substantial overall increase in fraction of variance explained (FVE) across different sites, shown in Fig.\,\ref{Figure2}B. 
A nonlinear decoder using only two best cells per site outperforms, on average, the best sparse linear decoder constructed from the entire population; nonlinear performance saturates quickly with the number of cells and peaks when decoding from local $\sim 8$-cell groups. An alternative way to compare decoding performance is to threshold the sequence of decoded movie frames (see SI~Fig.\,8 and SI Movie 1), thereby assigning each site to a decoded dark disc (``below threshold'') or to the bright background (``above threshold''). Decoded movie frames can then be compared to ground truth at each threshold using the receiver operator characteristic (ROC curve), shown for both decoders in Fig.\,\ref{Figure2}C. In this metric, nonlinear decoders also consistently outperformed linear ones. Excess nonlinear performance of between 15 and 20\% of FVE was maintained even when both decoders were trained on 10-disc stimulus and tested on stimuli with smaller number of discs (Fig.\,\ref{Figure2}D). Excess nonlinear performance was also observed when decoding from a cell mosaic of a single functional type (SI Fig.\,9) and on a repeat experiment (SI Fig.\,10).

A particularly striking feature of our results was the difficulty of the linear decoder to match the true (constant) luminance trace when no disc was crossing the corresponding site. Rat retinal ganglion cells are continuously active even when there are no coincident on-center luminance changes, with the activity likely resulting from stimulus changes in the surround, from long-lasting sustained responses to previous stimuli, from effective network coupling to cells that do experience varying input, or from true spontaneous excitation that would take place even in complete absence of stimuli~\cite{kuffler+al_1957,troy+lee_94,shlens+al_06,freeman+al_08}. Either way, activity of cells at constant local luminance presents a confound that is difficult for a generic linear mechanism to eliminate, which results in decoder fluctuations, or ``hallucinations,'' of sizable variance. To quantify this effect, we partitioned the luminance traces at every site into constant and fluctuating epochs by means of a simple threshold (see Methods), and examined decoding errors separately during both epochs. While the absolute error of the nonlinear decoder was smaller than that of linear in both epochs, the fractional difference was greatest during constant epochs, suggesting that nonlinear decoders might specifically be better at suppressing their responses to spontaneous-like neural activity (Fig.\,\ref{Figure2}E). We reasoned that this improvement comes, in part, from the  ability of the nonlinear method to recognize whether there are any on-site luminance fluctuations or not, from the spike trains alone. To test this idea, we trained linear and nonlinear classifiers, operating on identical inputs and with the same kernel parameters as the decoders, to best separate constant from fluctuating activity. Consistent with our expectations, nonlinear classifiers outperformed linear at every site, irrespectively of whether their input were the rasters of all local cells that contribute to the decoding, as shown in Fig.\,\ref{Figure2}F, or the raster of a single best cell at every site (SI~Fig.\,11).  

\begin{figure}
\centerline{\includegraphics{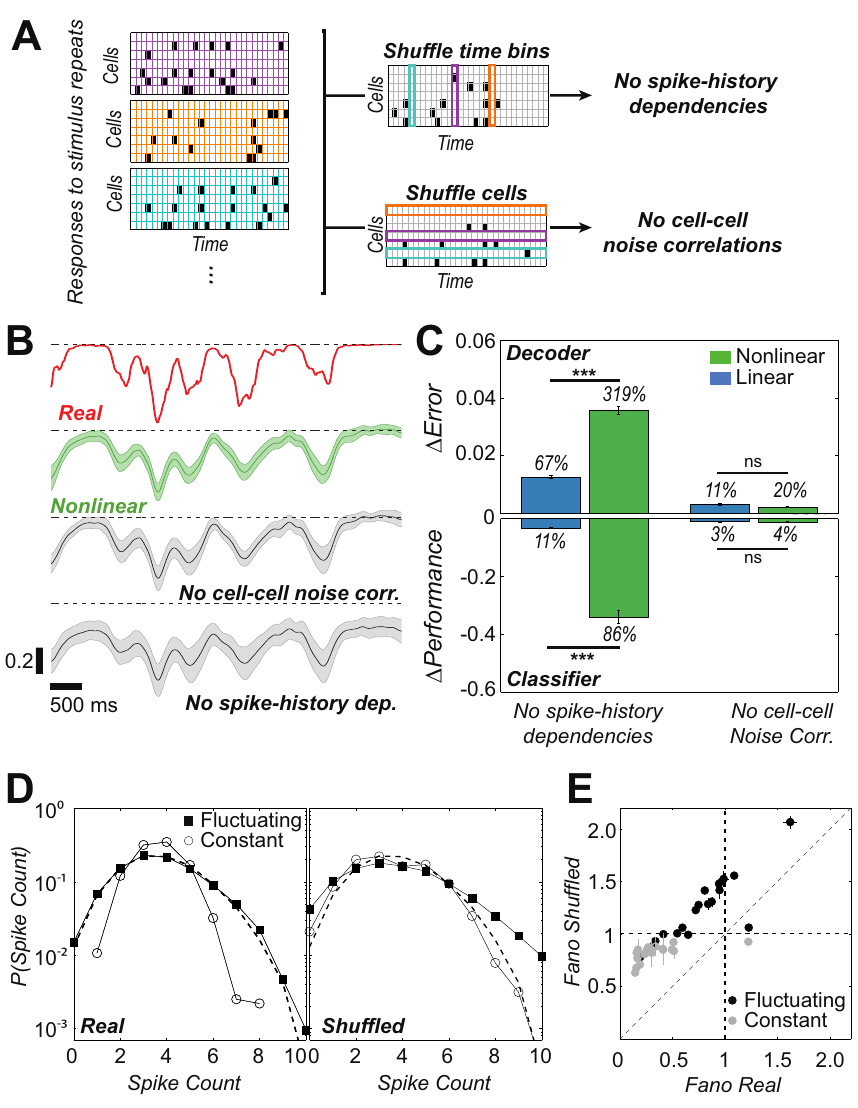}}
\caption{\textbf{Spike-history dependencies affect decoding performance.}
\textbf{A}: Shuffles of responses to repeated stimulus presentations remove different types of correlations, but preserve average locking to the stimulus (PSTH). 
\textbf{B}: A repeated stimulus fragment (red trace), nonlinear decoder predictions using real responses (green), and using responses without different types of correlations (gray); shown is the prediction mean $\pm$ SD over repeats.
\textbf{C}: Increase in decoding error (MSE) and decrease in classifier performance (F-score) when spike-history dependencies or noise correlations are removed (average $\pm$ SEM across sites); percentages report fractional differences relative to the original performance. 
\textbf{D}: Spike count distributions for a single example cell. Removing spike-history dependencies broadens the distributions, in particular in constant epochs. Dashed line = expectation for a fully randomized spike train with a matched firing rate. 
\textbf{E}: Fano factor of spike count distributions for spike trains with and without spike-history dependencies. Each point is a cell that contributes most to decoding at a particular site (when the same cell contributes to multiple sites, average $\pm$ SD across sites is shown).  
 }\label{Figure3}
\end{figure}

\subsection{Nonlinear decoders make use of spike-history dependencies in individual spike trains}
Next, we attempted to identify the statistics of spike trains that are necessary to explain the excess performance of nonlinear decoders. Our starting point was the following observation: the simplest nonlinear decoders that used a single best cell for each site, when interrogated with a test-set epoch of pure spontaneous activity (i.e., neural responses to a completely blank screen), yielded luminance traces with significantly smaller variance than their linear counterparts (SI~Fig.\,12). Since the only structure in spike trains during spontaneous activity is, by definition, due to ``noise correlations''---pairwise or higher-order dependencies between spikes within an individual spike train or across different spike trains---we hypothesized that certain noise correlations could be used by nonlinear decoders also during stimulus presentation to boost their decoding performance. 

To test this hypothesis, we made use of many identical repeats of a particular stimulus fragment embedded in our disc movie (these repeats were used neither for training nor testing). Using the same decoders as above, we decoded the original response rasters corresponding to the repeated fragment, as well as rasters in which we shuffled the spikes to remove spike-history dependencies, or to remove cell-cell noise correlations, as shown in Fig.\,\ref{Figure3}A; note that these manipulations left the firing rates of all cells intact. 
Figure~\ref{Figure3}B shows a stimulus reconstruction at an example site by the nonlinear decoder, for original rasters as well as rasters with removed spike-history dependencies or cell-cell noise correlations. Removing cell-cell noise correlations leads to a small increase in the variance of the reconstructions across stimulus repeats, with only marginal differences in the mean reconstructed trace, compared to decoding from intact rasters. Surprisingly, removing spike-history dependencies leads to much worse reconstructions, whose mean is strongly biased and variance increased; as a result, the dynamic range of the decoded trace is substantially lower compared to decoding from intact rasters. These observations are summarized across sites in Fig.\,\ref{Figure3}C, which shows the increase in decoding error and decrease in classifier performance when spike-history dependencies or cell-cell noise correlations are removed. Removal of cell-cell noise correlations leads to small increases in error, roughly of the same magnitude for both linear and nonlinear decoders; in contrast, while removal of spike-history dependencies leads to increases in error for both decoders, the effect is more than three-fold larger for the nonlinear decoder. Qualitatively similar conclusions hold for the classifiers trained to separate constant from fluctuating input epochs (Fig.\,\ref{Figure3}C), as well as for decoders and classifiers trained on the single best cell per site (SI~Fig.\,13).

Having established that spike-history dependencies are crucial to the performance of the nonlinear decoder, we looked at the detailed statistical structure of individual spike trains. For each neuron that best decoded the luminance trace at a specified site, we focused on $250\e{ms}$ (20 time bins) response sequences and constructed a distribution over the number of occupied time bins (``spike counts''), separately for epochs where the luminance trace was fluctuating or where it was constant. As shown in Fig.\,\ref{Figure3}D, these distributions differed significantly: the  count distribution was much tighter in constant epochs, while the mean firing rate between the epochs did not change much. During fluctuating-input epochs, observing more spikes in a $250\e{ms}$ window was more likely than at constant input, but---perhaps surprisingly---patterns with very low numbers of spikes (e.g., zero or one) were also more likely during fluctuating-input epochs. The count distribution at fluctuating light was very similar to binomial (and, at this temporal resolution, Poisson), while it was tighter at constant light. These changes could be summarized by a simple statistic, the Fano factor $F=$ (variance in spike count)$/$(mean spike count). When we removed spike-history dependencies, Fano factor increased for both distributions and they became harder to distinguish from each other. Figure~\ref{Figure3}E shows that this behavior was consistent across all sites, highlighting the very high regularity of neural spiking that resulted in sub-Poisson variance ($F$ substantially below 1) during epochs of constant luminance. 

Taken together, our results show that: \emph{(i),} spike-history dependencies within individual spike trains are crucial for nonlinear decoder performance; \emph{(ii),} these dependencies shape the distribution of spike counts on timescales relevant for decoding; \emph{(iii),} during constant local luminance, spiking activity is very regular (and statistically similar to true spontaneous activity, see SI~Fig.\,14); \emph{(iv),} a simple statistic, which summarizes the effects of spike-history dependencies in different epochs and their changes when the spike trains are shuffled, is the Fano factor. While this does not imply that kernelized decoders actually compute some version of a local estimate for the Fano factor (they could be sensitive to other statistics, e.g., the interspike interval distribution, which also differs substantially between the epochs, see SI~Fig.\,15), it is plausible that the underlying reason for nonlinear decoder performance is its ability to recognize high regularity of spiking during epochs of constant local luminance.

\subsection{A simple neural encoding model can recapitulate spike train statistics crucial for nonlinear decoding}
\begin{figure}
\centerline{\includegraphics{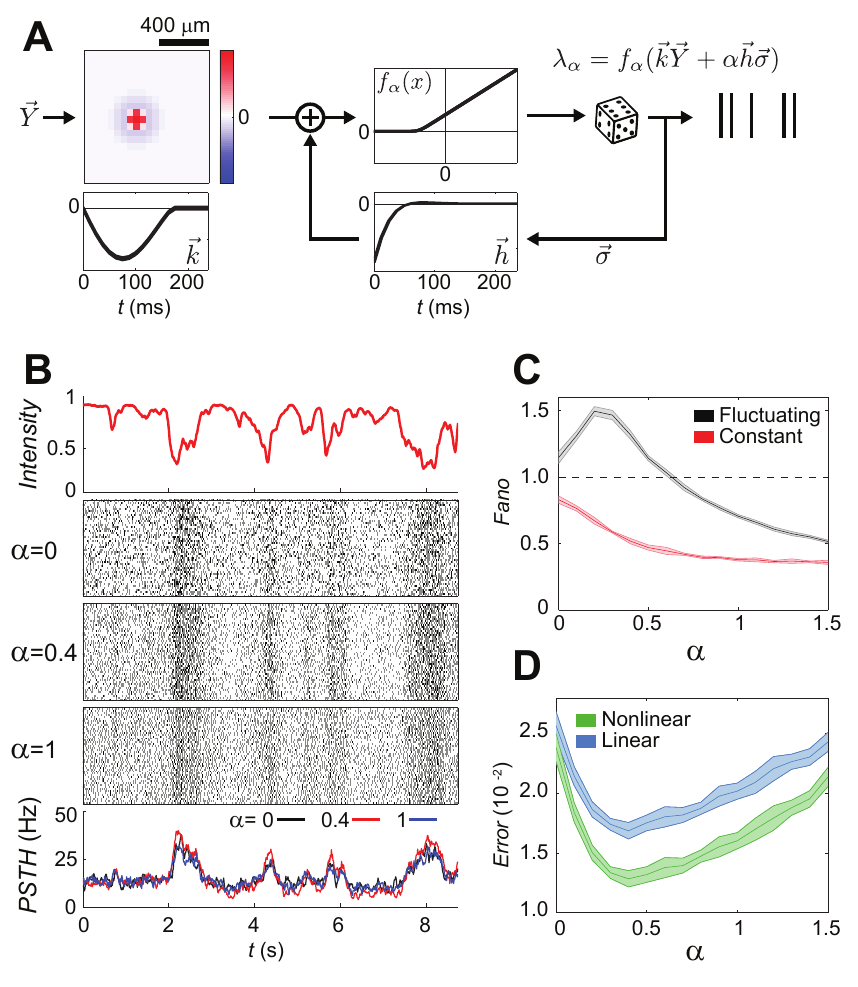}}
\caption{\textbf{Spike-history dependencies of intermediate strength facilitate nonlinear decoding in simple models of neural processing.}
\textbf{A}: Schematic of a single-cell Generalized Linear Model (see Methods). The neuron's sensitivity to the stimulus is determined by a radially symmetric difference-of-Gaussians spatial filter that has a monophasic timecourse ($\vec{k}$), and combines additively with the neuron's sensitivity to its own past spiking, given by filter $\vec{h}$ (with strong refractoriness followed by weak facilitation). Importantly, $\vec{h}$ shapes spike-history dependencies in the resulting spike trains. A nonlinear function $f(\cdot)$ (here, threshold-linear) of the combined sensitivities gives the neuron's instantaneous firing rate that can be used to generate individual spike train instances. Shapes, as well as the temporal and spatial scales of the filters, were realistic for our data. 
\textbf{B}: Example rasters (50 repeats) generated with the encoding model for a given intensity trace and different magnitudes ($\alpha$) of spiking history filter $\vec{h}$. The rasters are matched in PSTH (bottom) but differ in temporal noise correlations. 
\textbf{C}: Average Fano factor ($\pm$ SD) of the model as a function of $\alpha$ in fluctuating and constant epochs. 
\textbf{D}: Decoding error as a function of $\alpha$. Decoders are trained for each separate $\alpha$ and tested on withheld stimuli; shade = SD over 10 spike train realizations. 
}\label{Figure4}
\end{figure}

Can the observed spike-history dependencies, which enable successful nonlinear decoding, be generated by simple and generic neural encoding models? To address this question, we made use of generalized linear models (GLMs)~\cite{truccolo+al_05,pillow_07}, probabilistic functional models of spiking neurons that extend the paradigmatic linear-nonlinear (LN) framework by incorporating the recurrent feedback from neuron's past spiking, as schematized in Fig.\,\ref{Figure4}A.
Previously, GLMs have been successfully applied to responses of the mammalian retina \cite{pillow+al_2008,meytlis+al_2012} and in the cortex~\cite{truccolo+al_2010,lawhern+al_2010}, and also reproduced well the firing rates of cells recorded in our experiment on the repeated stimulus fragment (SI~Fig.\,16).

To link encoding models and decodability in a way that would generalize beyond the specifics of our dataset, we created the simplest stereotyped model cell, shown in Fig.\,\ref{Figure4}A.
Crucially, we parametrized the magnitude of the self-coupling filter with $\alpha$: $\alpha=0$ thus corresponded to a pure LN model, while increasing values of $\alpha$ made neural spike trains non-Poisson, progressively enforcing dependence on past spiking and consequently increasing the magnitude of the resulting temporal correlations.

With this model in hand, we generated a ``baseline'' raster  of repeated responses to a randomly moving disc stimulus at an initial value of $\alpha=1$, as shown in Fig.\,\ref{Figure4}B. The average firing rate was chosen to be the typical rate of our recorded ganglion cells. We then systematically changed the value of $\alpha$ and, for each value, refitted the nonlinearity to the baseline raster at $\alpha=1$ (see Methods). This procedure generated synthetic rasters that were matched in their peri-stimulus time histograms (PSTH) and stimulus preference, yet differed in the strength of spike-history dependencies. 

Following our previous analyses, we partitioned the luminance trace into constant and fluctuating epochs, and looked at the spiking statistics in $250\e{ms}$ (20 time bin) windows. Fano factor in constant epochs decreased as a function of $\alpha$ and dropped substantially below 1; in contrast, when on-center luminance was fluctuating, Fano factor behaved non-monotonically (Fig.\,\ref{Figure4}C). In line with expectations and behavior observed in our data, Fano factor at constant luminance was always below Fano factor at fluctuating luminance. Having ensured that the statistics of synthetic rasters qualitatively agreed with the data for the range of $\alpha$ we examined, we asked about the performance of linear and nonlinear decoders, trained and tested at different values of $\alpha$. Figure~\ref{Figure4}D plots the decoding error as a function of $\alpha$. Overall, the error levels are in range of those observed for real data (cf. Fig.\,\ref{Figure1}H), with nonlinear decoders outperforming linear by $\sim 10-30\%$. Interestingly, the minimal error for both decoders is achieved at an intermediate value of $\alpha^*\approx 0.4$, which also corresponds to the point where nonlinear decoders maximally outperform their linear counterparts. At $\alpha=0$, where the encoding models are effectively LN neurons, the decoders differ only marginally in performance (analogous results hold for the classifiers, see SI~Fig.\,17). 

In sum, for a generic class of encoding models that are widely applicable to both peripheral as well as central neural processing, there exists a non-trivial strength of spike-history dependence that facilitates stimulus reconstruction, especially with nonlinear readout. Intuitively, the existence of optimal $\alpha^*>0$ can be explained as a trade-off between ensuring regularity of spiking during constant epochs, which the nonlinear decoder can make use of, while not impeding stimulus encoding during fluctuating epochs; during these epochs, stimulus-driven term should dominate over sensitivity to past spiking, otherwise excessive dependence on spiking history (e.g., $\alpha \geq 1$ in Fig.\,\ref{Figure4}B) could perturb reliable locking to the stimulus.  
\section{Discussion}
Insights from decoding provide crucial constraints for theoretical models of neural codes. A large body of work dissects  nonlinearities in stimulus processing, from nonlinear summation in the receptive field or during adaptation, to essential spike generation nonlinearities. Consequently, one would expect  nonlinear decoding to outperform linear, but reports to that effect are surprisingly scarce~\cite{warland+al_1997,fernandes+al_2010}. In theory the results of a nonlinear encoding process can be linearly decodable~\cite{bialek+zee_1990,rad+paninski_2011}, yet whether this is true of real neurons under rich stimulation is still unclear. Another fundamental question concerns the stability of decoding transformations, which has recently received renewed attention  in the context of efficient coding~\cite{boerlin+deneve_2011,boerlin+al_2013,deneve+chalk_16}. Approaching this question empirically requires us to first construct high-quality decoders for complete stimulus movies---conceptually, doing the inverse of the state-of-the-art encoding models~\cite{pillow+al_2008}---which remains an open challenge. Finally, a number of studies, both theoretical~\cite{averbeck+al_2006} and data-driven~\cite{tkacik+al_2014,pillow+al_2008,meytlis+al_2012,schneidman+al_2006,ecker+al_2010,granot+al_2013}, focused on correlations in neural activity, especially those due to spike-history dependence and network circuitry (``noise correlations''); here, decoding provides a way to quantitatively ask about the functional contribution of such correlations to stimulus reconstruction.

We used large-scale linear and kernelized (nonlinear) regressions to directly decode a complex stimulus movie from the output of many simultaneously recorded retinal ganglion cells. Importantly, we did not use any prior knowledge of recorded cells' properties (e.g., their types or receptive fields), or any prior knowledge of the stimulus structure, to carry out the decoding; as a result, our decoding filters could, at least in principle, be used to decode any stimulus. A combination of sparse prior over decoding filter coefficients and a high-dimensional stimulus revealed a surprisingly local and stereotyped manner in which the retinal code could be read out. This is in stark contrast to previous work using simple stimuli where the readout was distributed and the resulting decoding filters had no general interpretation~\cite{marre+al_2015}. While our filters and consequently the ``decoding fields'' were recovered under a particular stimulus class and thus nominally depend on stimulus statistics, it is interesting to speculate whether the retina could adaptively change its encoding properties so as to keep the decoding representations constant, as has recently been suggested~\cite{schwartz+al_2012,marre+al_2015,frechette+al_2005}. Similarity between decoding and receptive fields and generalization to stimuli with different number of discs provide limited circumstantial support for this idea, but a definite answer can only emerge from dedicated experiments that specifically test the stability of decoders under rich stimuli with different statistical structure.

The performance of linear decoders was further improved by using nonlinear decoding. 
The improvement was significant, systematic, and reproducible: we observed it at nearly all sites, irrespectively of how many relevant cells we decoded from, when decoding from all recorded cells jointly or a mosaic of a single type, and also in a repeat experiment. Such an improvement is nontrivial, because the increased expressive power of kernelized methods comes at a cost of potentially overfitting models to data; this was evident also in our failed first attempt to apply nonlinear decoding to the whole recorded population, instead of only to the relevant cells selected by sparse linear decoder at every site. The performance improvement depended crucially on the spike-history dependence in individual spike trains but only slightly on cell-cell noise correlations~(cf.~\cite{pillow+al_2008,meytlis+al_2012}). 

What are the methodological advances presented in our work? First, the use of sparse linear and kernelized regression, as described here, should provide a tractable way of studying how rich signals are represented in other parts of the brain without making explicit assumptions about the encoding process, thereby providing a complementary, decoder-centric alternative to Bayes inversion of probabilistic encoding models. Second, even though the inner workings of kernelized methods are notoriously difficult to interpret intuitively, our analysis suggests that controlled manipulations of spike train statistics can provide valuable insights into which spike train features matter for decoding and which do not. Finally, we provide a preliminary account of how decoding of full complex movies can shed light on the functional contributions of different cell types to stimulus representation, by decoding from individual mosaics or from their combinations, and comparing the performance to that of a complete population.

What are the general implications of our results? The high-dimensional nature of our stimulus forced us to decode the movie ``pixel-by-pixel,'' rather than trying to decode its compact representation. This, in turn, focused our attention on the intermittent nature of signals to be decoded: at any given site, the luminance trace switched between epochs where nothing changed locally, and periods where the trace was fluctuating in time. Such intermittency is common to many natural stimuli across different sensory modalities \cite{hyvarinen_2009,nelken_99}, and therefore must shape the way in which sensory information is encoded \cite{Vinje+Gallant_2000, Froudarakis+al_2014, Baudot+al_2013}. From the decoding perspective, it can, however, also pose a serious challenge: since neurons might be similarly active irrespective of whether the stimulus fluctuates locally or not, a downstream processing layer would have to suppress ``hallucinations'' in response to upstream network-driven or spontaneous activity. We proposed a simple mechanism to that effect using history dependence of neural spiking: because neuronal encoding is nonlinear, the effect of spike-history dependence on neural firing substantially differs between epochs in which the neuron also experiences a strong stimulus drive and epochs in which it does not. In such situations, nonlinear methods can discriminate between a true stimulus fluctuation and spontaneous-like firing from statistical structure intrinsic to individual spike trains, even when the mean firing rate doesn't change appreciably between different epochs. This mechanism is not specific to the retina, and may well apply in other systems that display both stimulus-evoked and spontaneous activity.

\section{Materials and Methods}
\subsection{Data} 
Retinal tissue was obtained from adult (8 weeks old) male Long-Evans rat (Rattus norvegicus) and continuously perfused with Ames Solution (Sigma-Aldrich) and maintained at 32 $^{\circ}$C. Ganglion cell spikes were recorded extracellularly from a multi-electrode array with 252 electrodes spaced 60 $\mu$m apart (custom fabrication by Innovative Micro Technologies, Santa Barbara, CA). Experiments were performed in accordance with institutional animal care standards. 
The microelectrode covered a total retinal area of $\sim$ 1~mm$^2$. For the rat this corresponds to 16-17 degrees of visual angle \cite{Hughes_1979}. The spike sorting was performed with an in-house method based on \cite{mare+al_2012}.

\subsection{Visual Stimulus}
The stimulus movie consisted of randomly moving dark discs ($r$= 100~$\mu$m) against a bright background (100\% contrast, 2 $\cdot$ 10$^{12}$ photons/cm$^2$/s). The discs followed mutually avoiding trajectories generated through an Ornstein-Uhlenbeck process. The movie was divided in segments of 1, 2, 4 and 10 discs, each 675~s long. Segments with increasing number of discs were presented sequentially and in total 3 segments of each type were shown, amounting to a total experiment time of 135~min. Each segment was regularly interspersed with 18 short (7.5~s) clips of repeated stimulus: in sum, 54 repeated clips were shown for each stimulus with different number of discs.  The stimulus was convolved with a bank of 400 spatial symmetric gaussian filters ($\sigma$=66.67 $\mu$m) placed in a regular 20x20 grid to produce local luminance traces. The filter normalization ensures the resulting traces are bounded in (0,1). The width of the filters was selected in preliminary tests to optimize decoding performance. The movie stimulus was shown at a refresh rate of 80 Hz. The response spike trains were binned accordingly in bins of 12.5 ms, and time aligned to the stimulus. The spatio-temporal receptive fields of the retinal ganglion cells were obtained through reverse correlation to a flickering checkerboard stimulus. The checkerboard was constructed from squares of 130~$\mu$m that were randomly selected to be black or white at a rate of 40 Hz. Retinal spontaneous activity was recorded in full darkness (blackout condition) for 2.5~min.

\subsection{Linear decoder}
Let $\vec{y}$ be a one-dimensional stimulus trace of length $N$ time bins. In the linear decoding framework we assume that an estimate of the stimulus $\hat{\vec{y}}$ can be obtained from the neural response $\Sigma$ as
$\hat{\vec{y}}=\Sigma\cdot\vec{L}$, where $\vec{L}$ is a linear filter. In this formulation, the response of the retina is represented by the matrix $\Sigma \in \mathbb{R}^{N\times(C\times\Delta T +1)}$, where $C$ is the number of cells and $\Delta T$ the size in bins of the time window we associate with a single point in $\vec{y}$ (for all analyses $\Delta T=61$ corresponding to a window stretching from -375 ms to 375 ms around the time bin of interest). The extra dimension is a column of ones to account for the bias term in the decoding. Thus, the decoding filter $\vec{L}$ is structured as 
$
\vec{L}=[L_0 \vec{L}_1 \vec{L}_2 \vec{L}_3 \dots \vec{L}_C],
$
where $\vec{L}_i$ is the filter corresponding to cell $i$ and $L_0$ is the bias term. We learned the filters $\vec{L}$ by minimizing the square error function with L1-regularization
\[
\chi^2=\frac{1}{N}(\hat{\vec{y}}-\vec{y})^{\top}(\hat{\vec{y}}-\vec{y})+\lambda \|\vec{L}\|_1.
\]
 To solve the minimization problem computationally we made use of the Lasso algorithm with the routines by Kim et al.~\cite{Kim+al_2007}. Data was divided into training and testing sets (4.9$\cdot$10$^{4}$ training points, 2.3$\cdot$10$^{4}$ testing points). The filters were obtained from the training set and all measures of performance refer to the testing set. Regularization parameter $\lambda$ was chosen through 2-fold cross-validation on the training set.  The regularization term ensures the sparsity of the filters. Due to this sparsity some cells have negligible filter norms and therefore do not contribute to the decoding. This allows us to establish a hierarchy of cells by sorting them according to their filter norm $\|\vec{L}_i\|_1$. ``Single-best cell'' for every site refers to the cell with the largest norm. ``Contributing cells'' are the subset of cells with largest norm that jointly account for at least half of the total filter norm $\sum_i\|\vec{L}_i\|_1 $. 

\subsection{Nonlinear decoder}
If instead of L1-regularization we enforce L2-regularization, the linear decoding filters can be obtained analytically through the normal equation
\[
\vec{L}=\Sigma^{\top}(\Sigma\Sigma^{\top}+\lambda I)^{-1}\vec{y}.
\]
 Thus, an estimate of the stimulus for some new data $\hat{\Sigma}$ is given by
\[
\hat{\vec{y}}=\hat{\Sigma}\cdot\vec{L}=\hat{\Sigma}\Sigma^{\top}(\Sigma\Sigma^{\top}+\lambda I)^{-1}\vec{y}.
\]
Since this expression only depends on products of spike trains, we can make use of the kernel trick and substitute the usual scalar product by some appropriate nonlinear function $k$ of the spike trains. In this way, we can express our nonlinear decoding problem as 
\begin{eqnarray*}
\hat{\vec{y}}&=&\kappa^{\top}(K+\lambda I)^{-1}\vec{y},\\
\kappa_{ij}&=&k(\hat{\vec{\sigma}}_i,\vec{\sigma}_j),\\
K_{ij}&=&k(\vec{\sigma}_i,\vec{\sigma}_j),
\end{eqnarray*}
where $\vec{\sigma}_i^\top\in\mathbb{R}^{1\times(C\times\Delta T +1)}$ is the $i$th row of matrix $\Sigma$. This is known as Kernel Ridge Regression ~\cite{Lampert_2009,Bishop_2006}. For our analyses we have used the Gaussian kernel 
\[
k(\vec{\sigma}_i,\vec{\sigma}_j)=\exp \big( -\frac{1}{2s^2}\|\vec{\sigma}_i-\vec{\sigma}_j \|^2_2 \big).
\]
Before computing the kernel, it is customary to turn the spike trains into smooth traces for the sake of performance \cite{Park2013}. We convolved our spike trains with a Gaussian filter of 3 time bins width. The data was divided into training and testing sets (9.8$\cdot$10$^{3}$ training points, 2.3$\cdot$10$^{4}$ testing points). The parameters $s$ and $\lambda$ were obtained through joint 3-fold cross-validation on the training set. The performance of the nonlinear decoder depends on the set of cells considered. Contrary to the linear case where L1-regularization can effectively silence cells by setting their filters to zero, this nonlinear framework cannot ignore cells in a similar way. Therefore, including in the analysis non-informative cells can decrease the generalization performance of the decoder.  To determine the best subset of cells for decoding we took advantage of the hierarchy of cells established by the linear L1-regularized decoding. We trained nonlinear decoders with progressively more cells (best cell, best two cells, etc.) and selected the subset of minimum decoding error on the training set (SI~Fig.~7).  Effectively, we jointly cross-validated the three parameters $s$, $\lambda$, and the subset size.

\subsection{Classifiers}
For classification purposes we assign each time bin to one of two classes: ``fluctuating'' or ``constant''. ``Fluctuating'' corresponds to discs moving over the site of interest and decreasing the light intensity in that site, while ``constant'' refers to the constant illumination of the site when no discs are present. To label the time bins we use a simple cut-off criterion plus two further correcting steps to account for retinal adaptation effects. First we label as ``fluctuating'' every bin with stimulus intensity less than 0.99. Then we apply these corrections: \emph{i)} Every identified ``constant'' segment shorter than 30 bins (375 ms) is relabelled as ``fluctuating,'' and \emph{ii)} The first 30 bins following a ``fluctuating'' segment are also labelled ``fluctuating.'' In this way the stimulus at each site is divided in segments of fluctuating and constant intensity. We train both linear and nonlinear Support Vector Machine (SVM) classifiers to determine, from the spike train response, whether a given time bin is labelled as ``constant'' or ``fluctuating''. Similarly to the decoding framework, to classify a given bin we consider a time window of $\Delta T=61$ bins around it in the response. For the nonlinear SVM we use the same gaussian kernel as in nonlinear decoding and the parameter values obtained when training the decoder. Note that this is not the optimal nonlinear classifier but allows us to evaluate the classifying power of the decoding kernel. 
 
\subsection{Measures of Performance}
Given a stimulus intensity trace $\vec{y}$ and the corresponding decoding prediction $\hat{\vec{y}}$ we define the decoding error as the Mean Squared Error 
$
\mathrm{MSE}=N^{-1}(\hat{\vec{y}}-\vec{y})^{\top}(\hat{\vec{y}}-\vec{y}).
$
We also make use of the related Fraction of Variance Explained defined as 
$
\mathrm{FVE}=1-(\mathrm{MSE}/\mathrm{Var}(y)).
$

To measure decoding performance from the fully decoded movie we build Receiver Operating Curves (ROC). We threshold the decoded intensity trace at each site. If intensity is below threshold, the presence of a disc in the site is predicted. By comparing the prediction to the original stimulus frames as function of the threshold we can evaluate the performance of the decoder as a balance between the True Positive (TP) and False Positive (FP) rates
\[
\mathrm{TPR}=\frac{\mathrm{TP}}{\mathrm{TP}+\mathrm{FN}},\quad
\mathrm{FPR}=\frac{\mathrm{FP}}{\mathrm{FP}+\mathrm{TN}}.
\]

To assess the performance of the SVM classifiers we use the $F_1$-score measure defined as 
\[
F_1=2\frac{P R}{P+R},
\]
where $P$ is the Precision and $R$ the Recall given by
\[
P=\frac{\mathrm{TP}}{\mathrm{TP}+\mathrm{FP}},\quad
R=\frac{\mathrm{TP}}{\mathrm{TP}+\mathrm{FN}}.
\]
For the binary classification task, ``fluctuating'' is defined as the positive class. 

Unless otherwise stated, all of the statistical significance tests were performed with the Wilcoxon signed rank test. 

\subsection{ON/OFF ratio bias estimation} For each site $s$ we determine the set of available cells as those located less than 300 $\mu m$ from the site. We call $C_s$ the total number of available cells at site $s$. In general, $C_s$ is the sum of ON and OFF subtype cells, $C_s=C_s^{\mathrm{on}}+C_s^{\mathrm{off}}$. If, from the available cells at site $s$, we pick a random subset of size $N=N^{\mathrm{on}}+N^{\mathrm{off}}$, the probability of choosing $N^{\mathrm{off}}$ cells is given by the hypergeometric distribution (random draw without replacement) 
\[
p(N^{\mathrm{off}}|s,N)=\frac{\binom{C_s^{\mathrm{off}}}{N^{\mathrm{off}}}\binom{C_s-C_s^{\mathrm{off}}}{N-N^{\mathrm{off}}}}{\binom{C_s}{N}}.
\]
The average probability over all sites considered is
\[
p(N^{\mathrm{off}}|N)=\frac{1}{S}\sum_{s=1}^S p(N^{\mathrm{off}}|s,N).
\]

Separately, for each site $s$ we have established a hierarchy of cells from their decoding filter norms. Following the hierarchy we create decoding sets of different size $N$ (the best cell, the best two cells, etc) and we count the number of OFF type cells $N^{\mathrm{off}}$ in them. We summarize this information in the histogram $M(N^{\mathrm{off}},N)$ that counts the number of sites where the decoding set of size $N$ contains $N^{\mathrm{off}}$ OFF cells. With this histogram we obtain an empirical probability 
\[
p_{\mathrm{emp}}(N^{\mathrm{off}}|N)=\frac{M(N^{\mathrm{off}},N)}{S},
\]
that we can compare with $p(N^{\mathrm{off}}|N)$. In particular, the bias reported in SI Fig.\,5 is given by 
\[
100\cdot \frac{p_{\mathrm{emp}}(N^{\mathrm{off}}|N)-p(N^{\mathrm{off}}|N)}{p(N^{\mathrm{off}}|N)}.
\]
Only sites with $N^{\mathrm{off}},N^{\mathrm{on}}\geq 2$ were considered for the comparison (n=115).

\subsection{Encoding Model}
We build an encoding model for a single cell, based on the standard GLM type model proposed by Pillow et al \cite{pillow+al_2008}. The cell spikes stochastically through a Poisson process with a time-dependent firing rate $\lambda(t)$ given by
$
\lambda(t)=f_\alpha(\vec{k}\vec{Y(t)}+\alpha\vec{h}\vec{\sigma(t)})
$
where $\vec{k}$ is a spatio temporal filter acting on stimulus $\vec{Y}$ and $\vec{h}$ is a temporal filter of the past spike history of the cell represented by $\vec{\sigma}$. The function $f(x)$ is a rectifying nonlinearity of the log-exp form $f(x)=a \log(b\exp(x+c))$. The stimulus filter $\vec{k}$ factorizes into separate spatial and temporal filters. The spatial component is given by a balanced difference of gaussians, with widths $\sigma_c=35\mu m$ for the positive and $\sigma_s=100 \mu m$ for the negative part, providing a symmetrical center-surround type filter.
 The temporal part of the filter is given by a single negative lobe of a $\sin$-like function. The filter for the past spike history takes the form 
\[
h(t)=A\sin(t+\frac{\pi}{2})\exp(B(-t+\frac{\pi}{2})).
\]
This filter inhibits firing after a spike but, depending on the values of the parameters, it can have a positive lobe after the inhibitory part that tends to increase the firing rate. We consider a span of 250 ms (20 bins) for both the past history filter and the temporal part of the stimulus filter.  All elements of the filter are fixed except for the rectifying nonlinearity that is changed according to the value of $\alpha$. Initially, the parameters of the nonlinearity $f_{\alpha=1}(x)$   are adjusted to provide an average firing rate similar to that observed in real data. The $\alpha=1$ model is taken as the ground-truth and every time $\alpha$ changes, the nonlinearity $f_\alpha(x)$ is fitted anew by maximizing the likelihood on $\alpha=1$ rasters, in order to reproduce the firing rate trace (PSTH) as closely as possible to the PSTH generated by $\alpha=1$.
The model neuron is stimulated with real data and the intensity trace at the central site of its receptive field is the stimulus considered for decoding. The model has been implemented using the Nonlinear Input Model toolbox \cite{McFarland_2013}.

\begin{acknowledgments}
We thank Matthew Chalk, Cristina Savin, and Jonathan D Victor for helpful comments on the manuscript. We also thank Christoph Lampert for useful discussions on kernel methods.  This work was supported by ANR OPTIMA, the French State program Investissements d'Avenir managed by the Agence Nationale de la Recherche [LIFESENSES: ANR-10-LABX-65], by a EC grant from the Human Brain Project (CLAP) and NIH grant U01NS090501 to OM, the Austrian Research Foundation FWF P25651 to VBS and GT. VBS is partially supported by contract MEC, Spain (Grant No. AYA2013-48623-C2-2 and FEDER Funds). SD was supported by a PhD fellowship from the region Ile-de-France. The funders had no role in study design, data collection and analysis, decision to publish, or preparation of the manuscript.
\end{acknowledgments}

\clearpage
\section*{Supplementary Information}

\setcounter{figure}{0}
\makeatletter 
\renewcommand{\thefigure}{SI \@arabic\c@figure}
\makeatother

\begin{figure}[h]
\centerline{\includegraphics[width=0.28\textwidth]{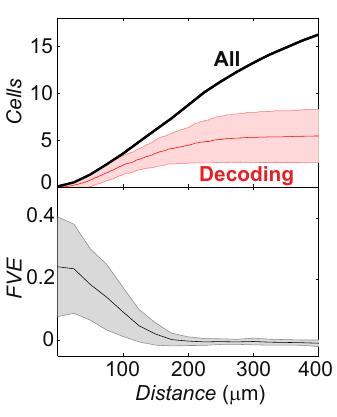}}
\caption{{\bf Decodable information is represented locally.} {\bf Top.} Average ($\pm$ SD) number of contributing cells (red) and all cells (black), as a function of distance of the cell's receptive field center to the site where the luminance trace is being decoded. {\bf Bottom.} Average ($\pm$ SD) single cell decoding performance as a function of distance to the site. Cells' responses contain no decodable information for sites that are $>200\e{\mu m}$ distant from their receptive field centers. Both analyses are done for the 10-disc stimulus.}
\end{figure}

\begin{figure*}[h]
\centerline{\includegraphics[width=0.55\textwidth]{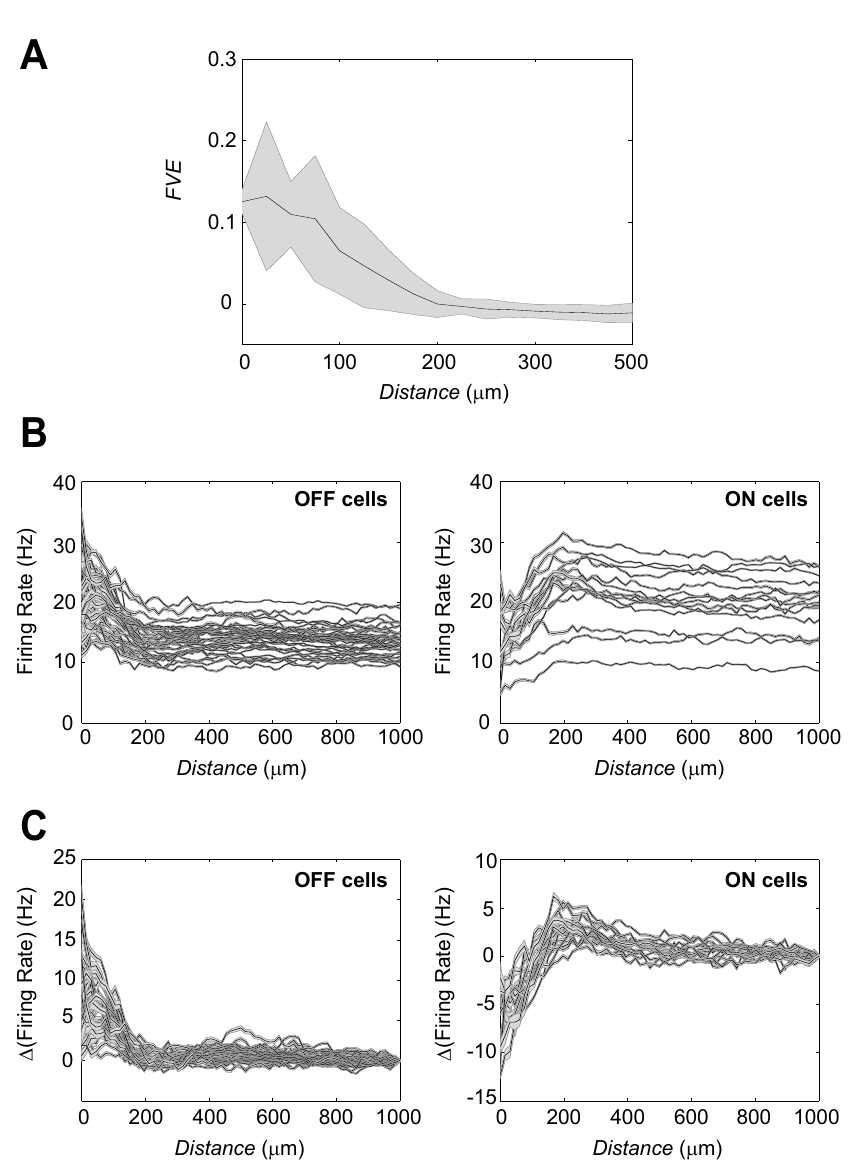}}
\caption{{\bf  Cells are continuously active, but their responses only contain decodable information about local luminance fluctuations.} The following analyses are carried out with a 1-disc stimulus. \textbf{A:} Average ($\pm$ SD) single cell decoding performance as a function of distance of the cell's receptive field center to the site where the luminance trace is being decoded. \textbf{B:} Firing rates of ON (N=14) and OFF (N=34) cells as a function of the distance to the single moving disc. Both types of cells exhibit basal firing rates $>10\e{Hz}$ when the disc is far away from their receptive fields. OFF cells increase their firing rate when the dark disc is less than 200 $\mu m$ away. ON cells decrease their firing in response to the dark disc and their firing rate peaks at the 200 $\mu m$ mark, probably corresponding with the stimulation of their surround by the dark disc. \textbf{C}: Same as in \textbf{B} but now the basal firing rate (measured at 1000 $\mu m$) has been subtracted for each cell to emphasize the stereotyped dynamics of the cells' activity. This analysis suggests that while cells are continuously active (even when the disc is far away and not stimulated by other discs, as in the case of SI Fig.\,1), that activity does not contain decodable information about the luminance fluctuations farther than $200\e{\mu m}$ from the receptive field center. In contrast, with simpler stimuli that stimulate retina more broadly (e.g., diffusively moving 1D bar), retinal ganglion cells encoded for the bar position in a distributed manner such that the stimulus could be decoded from multiple subsets of cells and even from cells whose receptive field centers were very distant from the bar position [Marre et al. 2015]. }
\end{figure*}

\begin{figure*}[h]
\centerline{\includegraphics[width=0.6\textwidth]{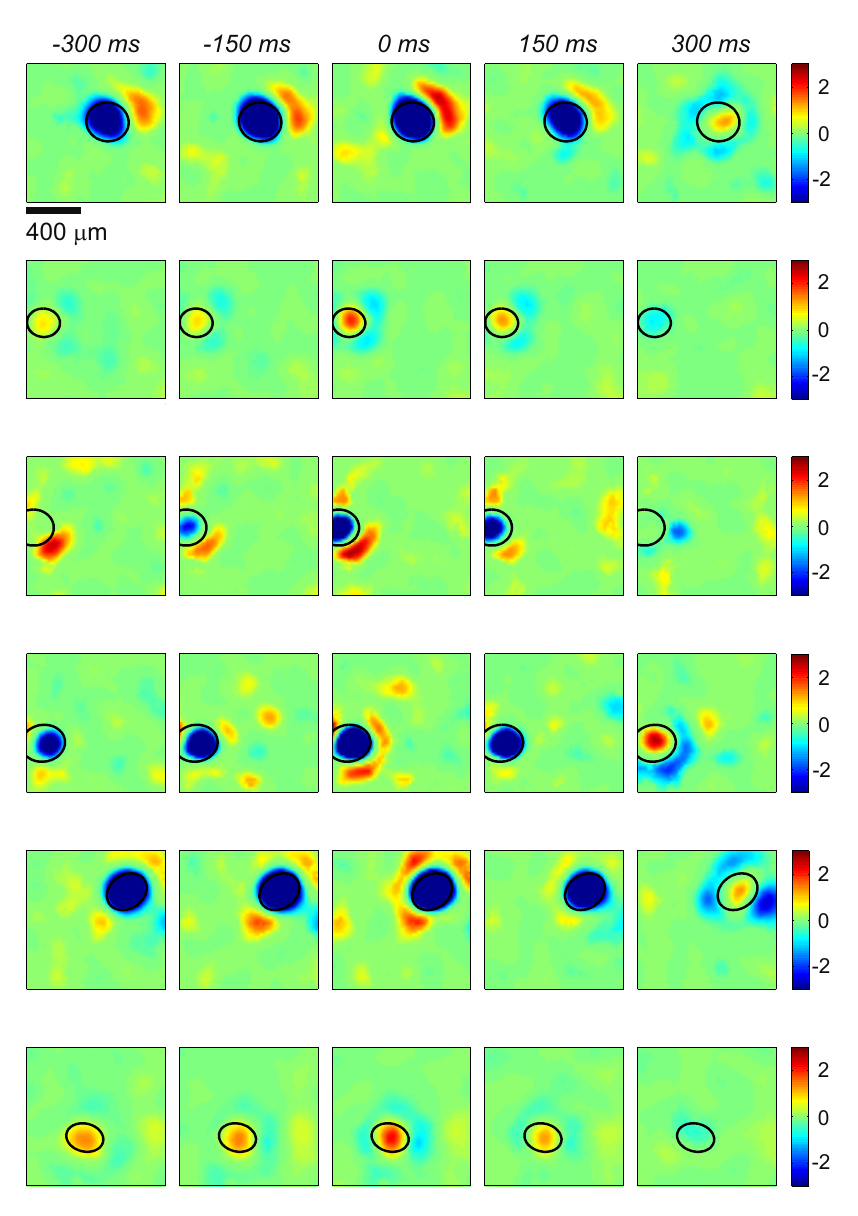}}
\caption{{\bf Examples of decoding fields for 6 different cells.} Each pixel corresponds to a site (of a 50 $\times$ 50 grid) and the color code represents the decoding filter of the cell at that particular site and time. The filters have been normalized such that the site of maximum variation has variance equal to 1. The white noise receptive field center of each cell is shown for reference (black ellipse).}
\end{figure*}

\begin{figure*}[h]
\centerline{\includegraphics[width=0.4\textwidth]{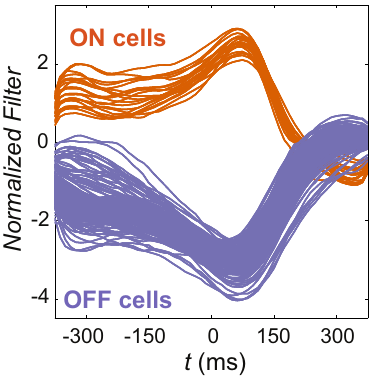}}
\caption{{\bf Decoding filters of best contributing cells have a stereotyped shape.} Decoding filters of the 1st and 2nd best contributing cells across sites, normalized to unit variance. The shape of the filters is very similar and differs primarily by a multiplicative scaling factor.
We could assume a universal temporal profile for all
cells at all sites, and perform the decoding by fitting a single
multiplicative scale parameter (with a sign, to account for ON/OFF differences) per cell per site, with less than
6\% drop in FVE on the 10-disc stimulus, compared to the model in the main text that makes no 
assumption about stereotyped filter shapes.}
\end{figure*}

\begin{figure*}[h]
\centerline{\includegraphics[width=0.4\textwidth]{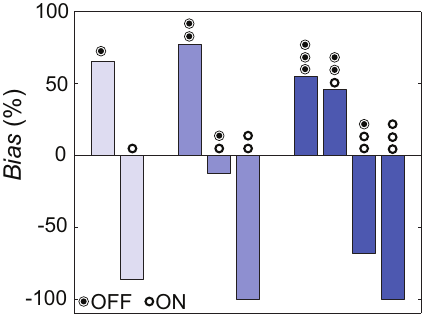}}
\caption{{\bf Decoding preferentially recruits OFF cells.} Bias in the ON/OFF cells ratio plotted separately for the single-, two- and three-best-cell decoding subsets for each site. By looking in detail at the contribution of ON vs OFF cells to stimulus reconstruction at every site we find
a clear bias for OFF cells relative to the prediction based on
random draws from the local ON/OFF composition (see Methods). This
OFF bias matched our expectation for optimally tracking dark
discs displayed in our experiments.}
\end{figure*}

\begin{figure*}[h]
\centerline{\includegraphics[width=0.5\textwidth]{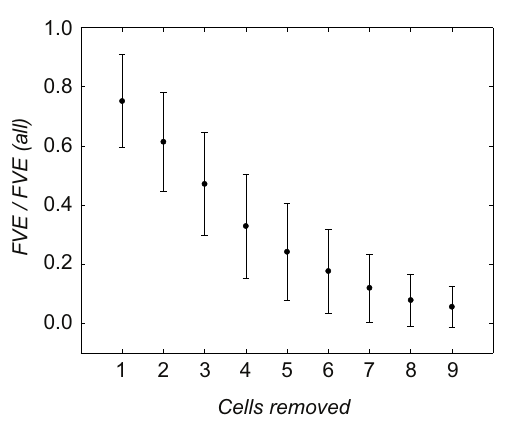}}
\caption{{\bf Redundancy of decodable information about local luminance traces.} Average fractional decrease in linear decoding performance across sites when progressively removing cells ($\pm$ SD). At each site cells are removed in order of importance, according to their decoding filter norm. We compare the performance when decoding with all available cells (FVE(all)) and when decoding without the first $N$ contributing cells (FVE). This is one way to estimate the redundancy in the population response. Removing 4-5 cells halves decoding performance, suggesting that the necessary information for linear decoding is contained in a small number of cells. This is in contrast with previous work [Marre et al. 2015], where we found that the information about the position of a moving bar was encoded in a highly redundant manner. In that work we were able to construct 5 disjoint subsets of cells (from 2 to 10 cells in size) from which the position of the bar could be decoded with low error. Together with SI Fig.\,2 this suggests that complex stimuli used here lead to much more local and less redundant responses that carry stimulus information (compared to e.g., diffusive bar motion), even though the retina is broadly active in both cases. }
\end{figure*}

\begin{figure*}[h]
\centerline{\includegraphics[width=0.9\textwidth]{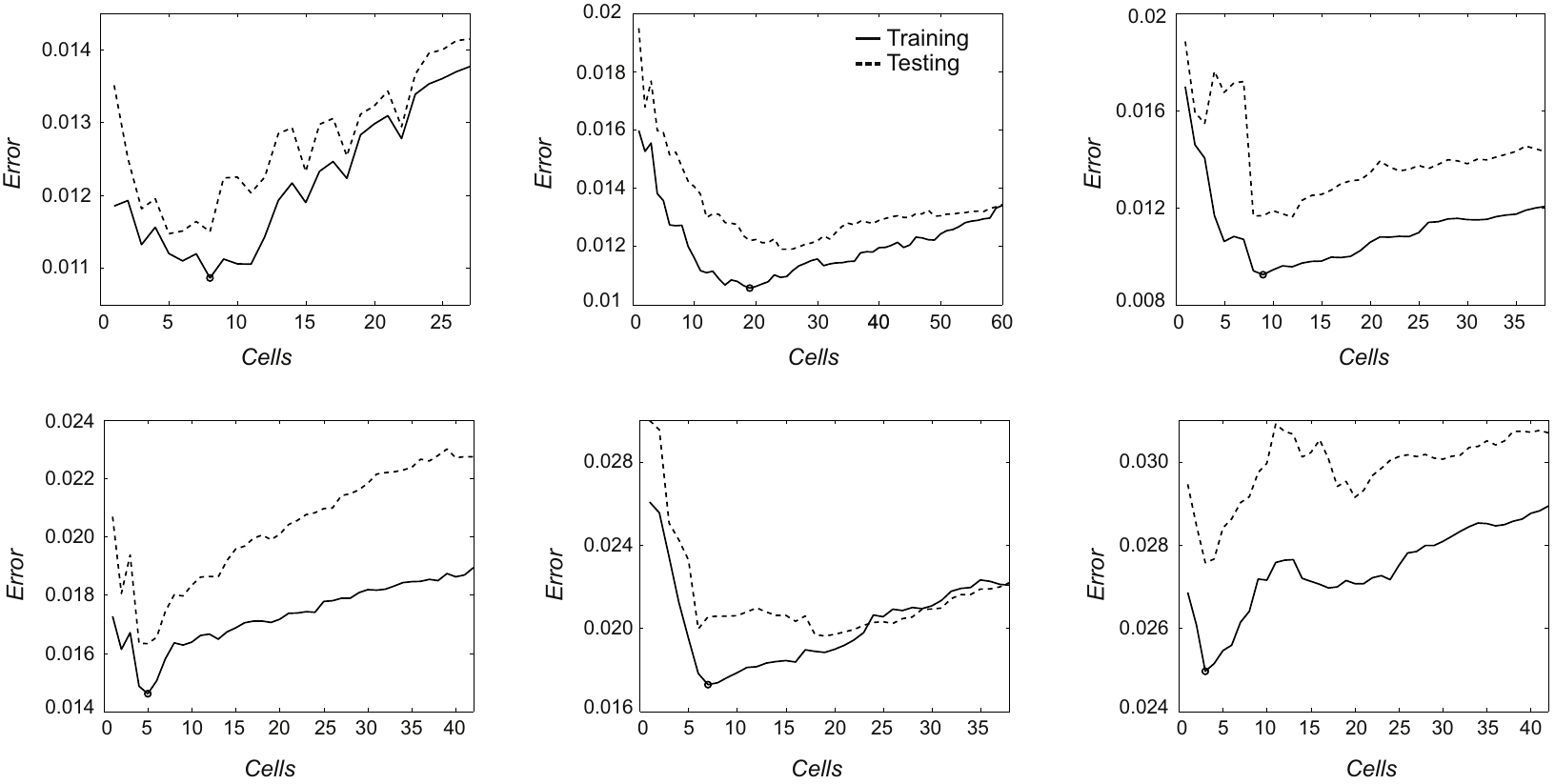}}
\caption{{\bf Choice of best subset of cells for nonlinear decoding.} Decoding error of the nonlinear decoder is plotted as a function of the number of cells considered for six different sites. Cells are ordered by the decreasing L1 norm of their linear filters (i.e., cell 1 is the best contributing cell, etc). The optimal subset (circle) is chosen through cross validation to minimize the error on the training set. The error of the nonlinear decoder on the test set is shown for comparison.}
\end{figure*}

\begin{figure*}[h]
\centerline{\includegraphics[width=0.6\textwidth]{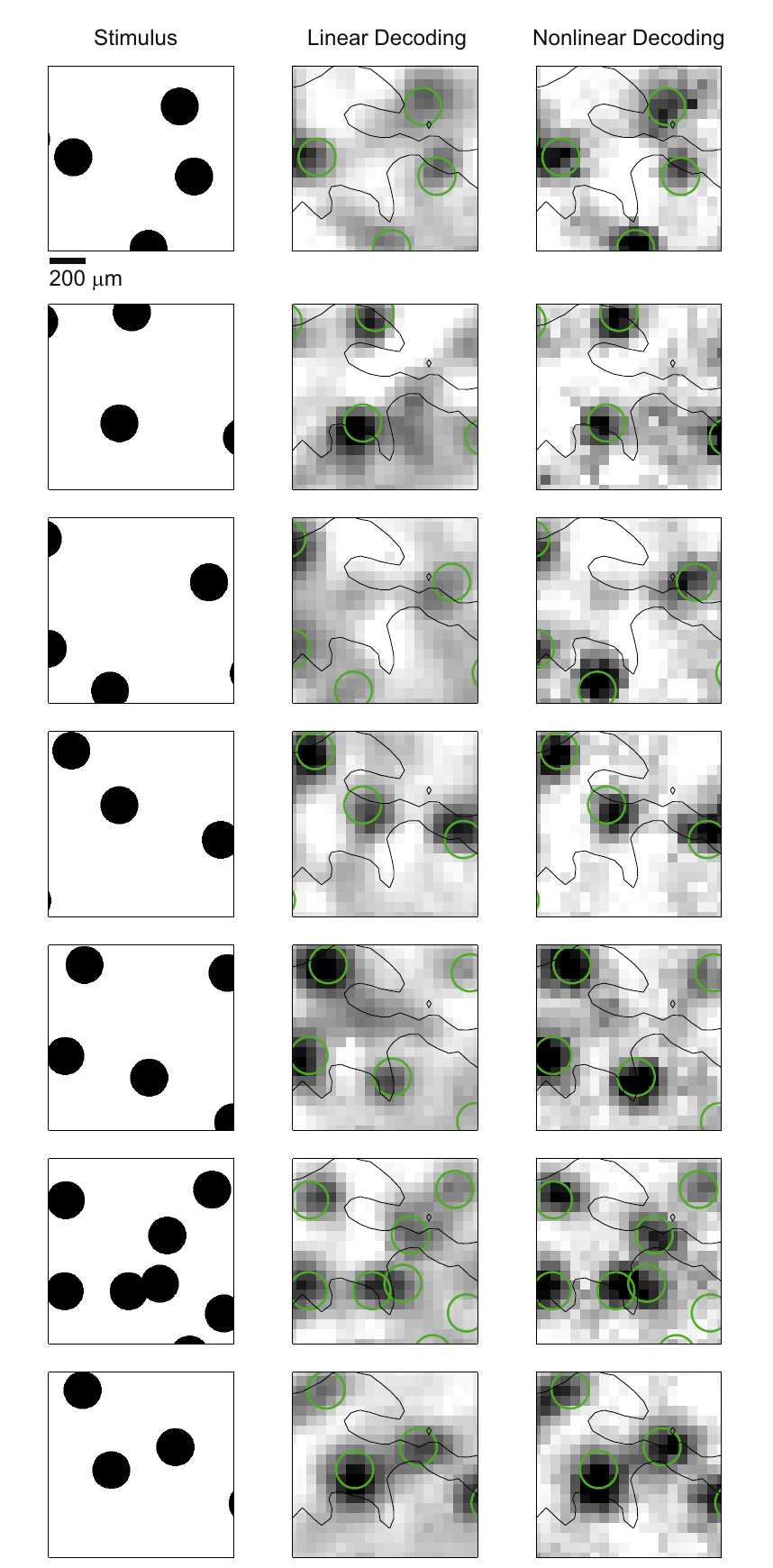}}
\caption{{\bf Examples of decoded movie frames with linear and nonlinear decoding.} Black contour marks the region of good cell coverage where linear decoding performs at $FVE>0.4$; green circles in decoded frames correspond to true positions of the discs.}
\end{figure*}

\begin{figure*}[h]
\centerline{\includegraphics[width=0.8\textwidth]{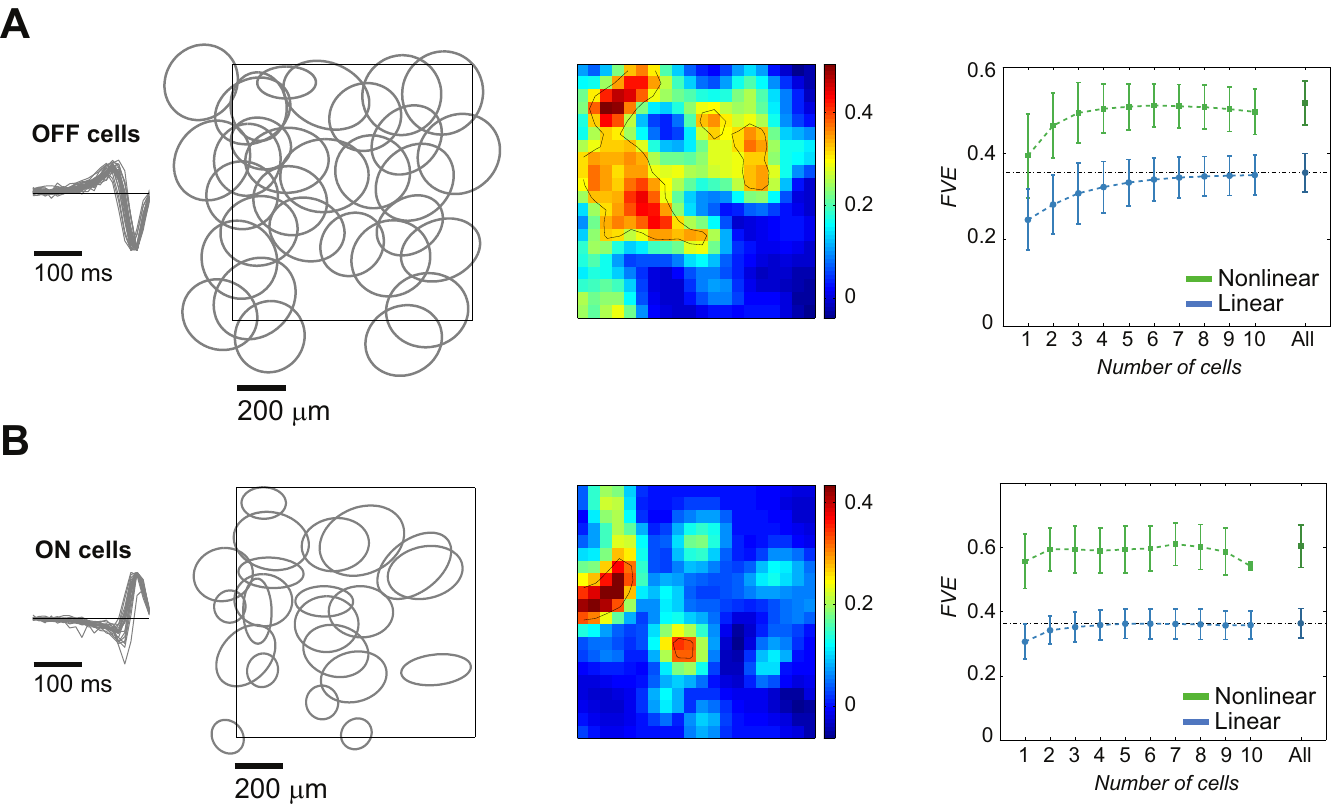}}
\caption{{\bf Decoding from single cell type mosaics.} \textbf{A}: OFF-cell mosaic (N=33). In the left-most panel temporal receptive field and spatial receptive field centers are shown. Center panel shows the performance of the linear decoders in space (measured as FVE). The contour lines mark the boundary FVE=0.3, and we only consider sites within this boundary to compute the average decoder
performance ($\pm$ SD across sites), achievable using increasing numbers of cells with highest L1
filter norm (right-most panel). For nonlinear decoding, ``All'' is the optimal subset that maximizes performance. \textbf{B}: ON-cell mosaic (N=22). Details equivalent to \textbf{A}. In both cases, nonlinear decoding substantially improves on linear.}
\end{figure*}

\begin{figure*}[h]
\centerline{\includegraphics[width=0.6\textwidth]{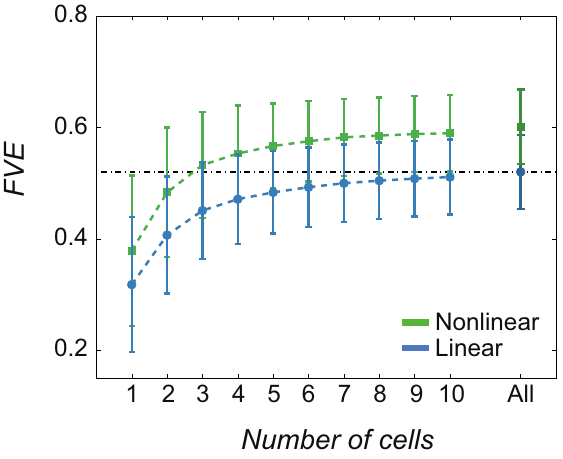}}
\caption{{\bf Decoding performance for a repeat experiment with a retina of a different rat}.  Average decoder performance ($\pm$ SD across sites), achievable using increasing number of cells with highest L1 filter norm. For nonlinear decoding, ``All'' is the optimal subset that maximizes performance. In the repeat experiment we isolated 64 retinal ganglion cells and identified 125 sites where linear decoding performed at FVE$>$0.4. }
\end{figure*}

\clearpage

\begin{figure*}[h]
\centerline{\includegraphics[width=0.5\textwidth]{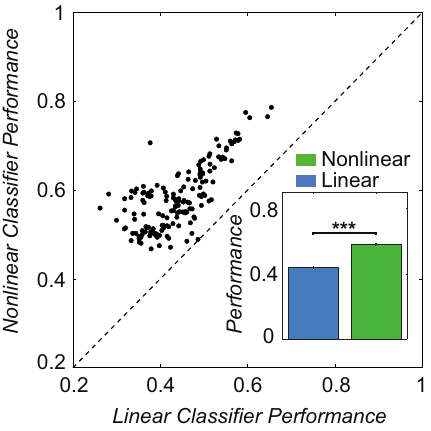}}
\caption{{\bf Nonlinear classifiers outperform linear on single cell responses.} Performance (F-score) of linear and nonlinear classifiers for each site when trained and tested from a single cell response (the best cell for each site). Average performance is shown in the inset ($\pm$ SEM) and the differences between linear and nonlinear are significant (p$<$0.001). }
\end{figure*}

\begin{figure*}[h]
\centerline{\includegraphics[width=0.6\textwidth]{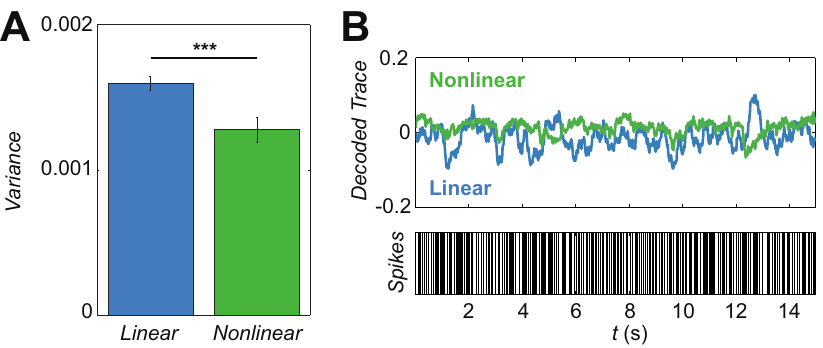}}
\caption{{\bf Nonlinear decoders predict more constant signal under blackout stimulation.} \textbf{A}: Variance of the single cell decoded traces from spontaneous activity (average across sites $\pm$ SEM). The decoders are trained on 10-discs stimulus and tested on the responses recorded during blackout condition (full darkness). Nonlinear decoders produce traces with significantly lower variance (p$<$0.001).  \textbf{B}: Example of mean-subtracted blackout decoded traces from a single cell spike train (bottom) with linear and nonlinear decoders.}
\end{figure*}

\begin{figure*}[h]
\centerline{\includegraphics[width=0.4\textwidth]{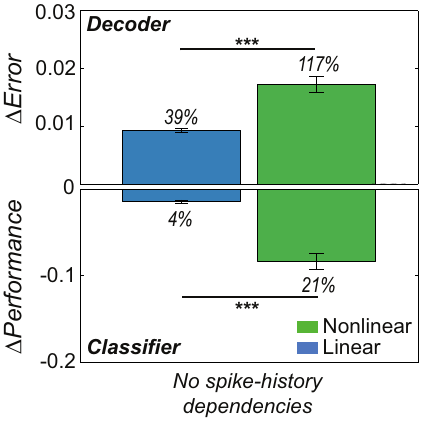}}
\caption{{\bf Nonlinear decoders (classifiers) rely on spike-history dependences when decoding (classifying)  single cell responses.} Changes in single cell decoders and classifiers performance when spike-history dependencies are removed. We show differences in average decoding error (MSE) for the decoder and differences in performance (F-score) for the classifier ($\pm$ SEM). The percentages shown stand for average fractional difference with respect to the original performance (before removing correlations). The differences are statistically significant in both cases (p$<$0.001).}
\end{figure*}

\begin{figure*}[h]
\centerline{\includegraphics[width=0.5\textwidth]{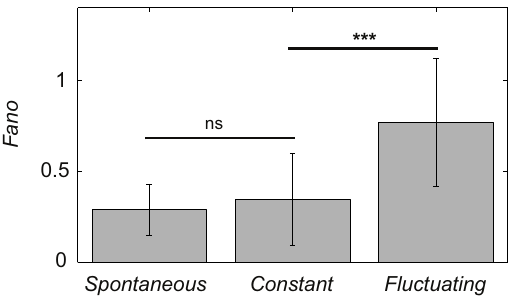}}
\caption{{\bf Fano factor is similar for responses at locally constant luminance and spontaneous activity, and differs for locally fluctuating luminance.} Fano factor, under different stimulus conditions, of the spike count distributions P(K) of the best cell for each site (average over sites $\pm$ SD). ``Spontaneous'' is the activity under blackout condition (no stimulus). The Fano factors of ``spontaneous'' and ``constant'' activities are not significantly different, pointing at similarities between these two responses. On the contrary,   both of them are clearly different from the activity under fluctuating stimulation (p$<$0.001).}
\end{figure*}

\begin{figure*}[h]
\centerline{\includegraphics[width=0.7\textwidth]{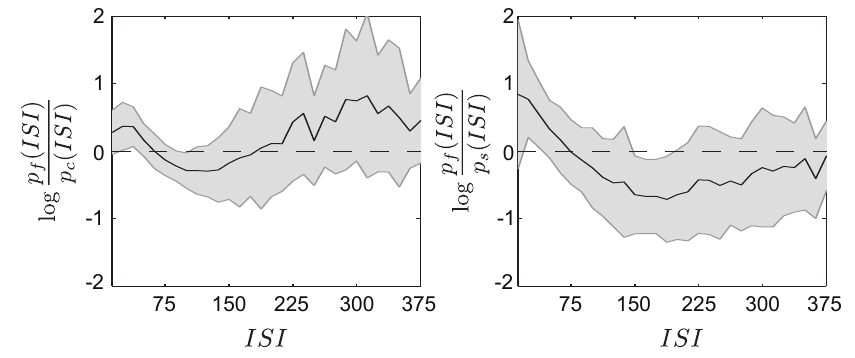}}
\caption{{\bf Interspike interval distributions differ at locally fluctuating luminance and locally constant luminance or spontaneous activity.} Logarithmic differences between the Inter-Spike-Interval (ISI) distributions under fluctuating [$p_f(ISI)$] and constant [$p_c(ISI)$] stimulus and between fluctuating and spontaneous activity [$p_s(ISI)$].  The distributions are computed for the single best cell at each site. The average across sites ($\pm$ SD) is shown. Similarly to the spike count distributions $P(K)$, the $ISI$ distributions show activity under constant stimulation to be more regular and dominated by $ISI$ between 75~ms and 175~ms. $ISI$ outside this range are more common during fluctuating stimulation.}
\end{figure*}

\begin{figure*}[h]
\centerline{\includegraphics[width=1.0\textwidth]{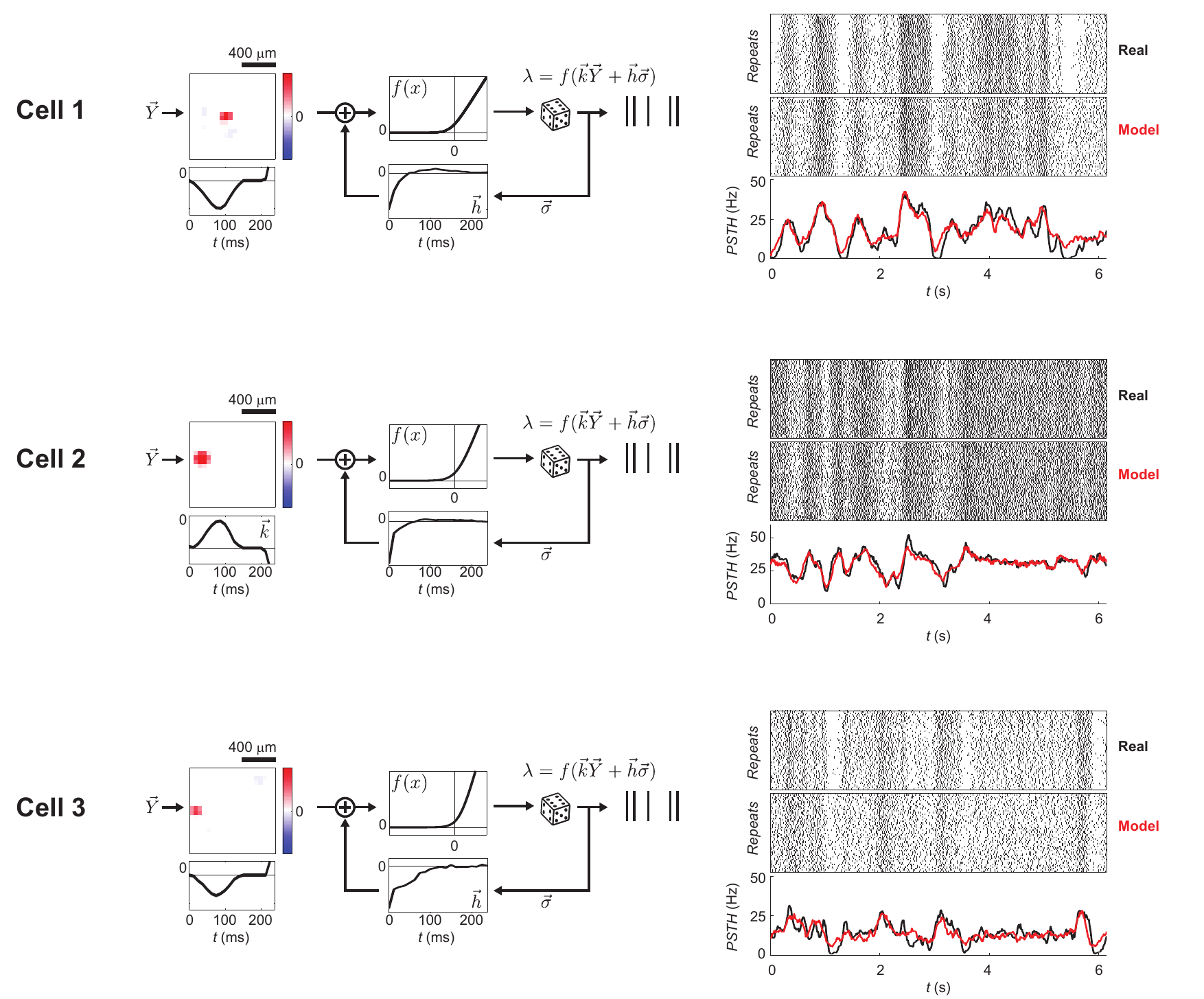}}
\caption{{\bf GLM models account well for the firing rates of cells recorded in the 10-disc experiment.} Three examples of GLM fits of real cells in our data set. On the left we show the fitted filters, nonlinearity, and spike history term that compose the model. On the right we show real and model generated repeated stimulus raster responses, and compare the real and predicted PSTH.  }
\end{figure*}

\begin{figure*}[h]
\centerline{\includegraphics[width=0.5\textwidth]{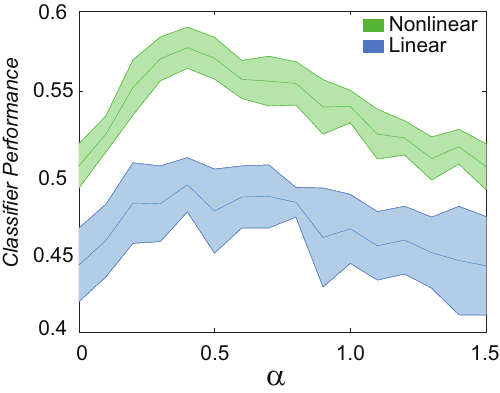}}
\caption{{\bf Classifier performance for constant vs fluctuating local luminance peaks at an intermediate value of spike-history dependencies.} Average classifier performance (F-score) as a function of $\alpha$ (see Fig. 4 in the main text). The error bars correspond to standard deviation over 10 different realizations of the spike trains generated from the model for each value of $\alpha$.}
\end{figure*}



\begin{thebibliography}{99}

\bibitem{spikesbook}
Rieke F, Warland D, de Ruyter van Steveninck RR, Bialek W (1997) Spikes: Exploring the Neural Code (MIT Press, Cambridge).

\bibitem{oram+al_1998}
Oram MW, Foldiak P, Perrett DI, Sengpiel F (1998) The `ideal homunculus': decoding neural population signals. \emph{Trends Neurosci} {\bf 21:} 259--65.

\bibitem{georgopoulos+al_1986}
Georgopoulos AP, Schwartz AB, Kettner RE (1986) Neuronal population coding of movement direction. \emph{Science} {\bf 233:} 1416--1419.

\bibitem{kay+al_2008}
Kay KN, Naselaris T, Prenger RJ, Gallant JL (2008) Identifying natural images from human brain activity. \emph{Nature} {\bf 452:} 352--5.

\bibitem{strong+al_1998}
Strong SP, Koberle R, de Ruyter van Steveninck RR, Bialek W (1998) Entropy and information in neural spike trains. \emph{Phys Rev Lett} {\bf 80:} 197.

\bibitem{archer+al_2013} 
Archer E, Park IM, Pillow JW (2013) Bayesian entropy estimation for binary spike train data using parametric prior knowledge. \emph{Advances Neural Info Proc Syst} {\bf 26:} 1700--1708.

\bibitem{tkacik+al_2014}
Tka\v{c}ik G et al. (2014) Searching for collective behavior in a large network of sensory neurons. \emph{PLOS Comput Biol} {\bf 10:} e1003408.

\bibitem{borst+theunissen_1999}
Borst A, Theunissen FE (1999) Information theory and neural coding. \emph{Nat Neurosci} {\bf 2:} 947--57.

\bibitem{quiroga+panzeribook}
Quiroga RQ, Panzeri S (2013) Decoding and information theory in neuroscience, pg. 139--163. In Principles of Neural Coding, Quiroga and Panezeri, eds. (CRC Press, Boca Raton, FL, USA).

\bibitem{Destexhe+Contreras_2006}
Destexhe A, Contreras D (2006) Neuronal computations with stochastic network states. \emph{Science} {\bf 314:} 85--90.

\bibitem{major+tank_04}
Major G, Tank DW (2004) Persistent neural activity: prevalence and mechanisms. \emph{Curr Opin Neurobiol} {\bf 14:} 675--684.

\bibitem{Ringach_2009}
Ringach DL (2009) Spontaneous and driven cortical activity: implications for computation. \emph{Curr Opin Neurobiol} {\bf 19:} 439--444.

\bibitem{tsodyks+al_99}
Tsodyks M, Kenet T, Grinvald A, Arieli A (1999) Linking spontaneous activity of single cortical neurons and the underlying functional architecture. \emph{Science} {\bf 286:} 1943--1946.

\bibitem{kuffler+al_1957}
Kuffler SW, Fitzhugh R, Barlow HB (1957) Maintained activity in the cat's retina in light and darkness. \emph{J Gen Physiol} {\bf 40:} 683--702.

\bibitem{troy+lee_94} 
Troy JB, Lee BB (1994) Steady discharges of macaque retinal ganglion cells. \emph{Vis Neurosci} {\bf 11:} 111-118. 

\bibitem{shlens+al_06}
Shlens J, Field GD, Gauthier JL, Grivich MI, Petrusca D, Sher A, Litke AM, Chichilnisky EJ (2006) The structure of multi-neuron firing patterns in primate retina. \emph{J Neurosci} {\bf 26:} 8254--8266.

\bibitem{freeman+al_08}
Freeman DK, Heine WF, Passaglia CL (2008) The maintained discharge of rat retinal ganglion cells. \emph{Vis Neurosci} {\bf 25:} 535--548.

\bibitem{bialek+al_1991}
Bialek W, Rieke F, de Ruyter van Steveninck RR, Warland D (1991) Reading a neural code. \emph{Science} {\bf 252:} 1854--7.

\bibitem{warland+al_1997}
Warland DK, Reinagel P, Meister M (1997) Decoding visual information from a population of retinal ganglion cells. \emph{J Neurophysiol} {\bf 78:} 2336--2350.

\bibitem{marre+al_2015}
Marre O et al. (2015) High accuracy decoding of a dynamical motion from a large retinal population. \emph{PLOS Comput Biol} {\bf 11:} e1004304.

\bibitem{schwartz+al_2012}
Schwartz G, Macke J, Amodei D, Tang H, Berry MJ 2nd (2012) Low error discrimination using a correlated population code. \emph{J Neurophysiol} {\bf 108:} 1069--88.

\bibitem{frechette+al_2005}
Frechette ES et al. (2005) Fidelity of the ensemble code for visual motion in primate retina. \emph{J Neurophysiol} {\bf 94:} 119--135.

\bibitem{pillow+al_2008}
Pillow JW et al. (2008) Spatio-temporal correlations and visual signaling in a complete neural population. \emph{Nature} {\bf 454:} 995--9.

\bibitem{meytlis+al_2012}
Meytlis M, Nichols Z, Nirenberg S (2012) Determining the role of correlated firing in large populations of neurons using white noise and natural scene stimuli. \emph{Vision Res} {\bf 70:} 44--53.

\bibitem{nichols+al_2013}
Nichols Z, Nirenberg S, Victor JD (2013) Interacting linear and nonlinear characteristics produce population coding asymmetries between ON and OFF cells in the retina. \emph{J Neurosci} {\bf 33:} 14958--14973. 

\bibitem{paiva+al_2009}
Paiva ARC, Park IM, Principe JC (2009) A reproducing kernel Hilbert space framework for spike train signal processing. \emph{Neural Comput} {\bf 21:} 424--449.

\bibitem{mare+al_2012}
Marre O et al. (2012) Mapping a complete neural population in the retina. \emph{J Neurosci} {\bf 32:} 14859--14873.

\bibitem{Park2013}
Memming Park I. et al. (2013) Kernel methods on spike train space for neuroscience: A tutorial. IEEE Signal Processing Magazine 30(4). 

\bibitem{truccolo+al_05}
Truccolo W, Eden UT, Fellows MR, Donoghue JP, Brown EN (2005) A point process framework for relating neural spiking activity to spiking history, neural ensemble, and extrinsic covariate effects. \emph{J Neurophysiol} {\bf 14:} 1074--1089.

\bibitem{pillow_07}
Pillow JW (2007) Likelihood-based approaches to modeling the neural code. In \emph{Bayesian Brain: Probabilistic Approaches to Neural Coding}, K Doya, S Shiii, A Pouget, R Rao eds, pg. 53--70. MIT Press (Cambridge, MA, USA).

\bibitem{truccolo+al_2010}
Truccolo W, Hochberg LR, Donoghue JP (2010) Collective dynamics in human and monkey sensorymotor cortex: predicting single neuron spikes. \emph{Nature Neurosci} {\bf 13:} 105--113.

\bibitem{lawhern+al_2010}
Lawhern V, Wu W, Hatsopoulos N, Paninski L (2010) Population decoding of motor cortical activity using a generalized linear model with hidden states. \emph{J Neurosci Methods} {\bf 189:} 267--280.

\bibitem{fernandes+al_2010}
Fernandes NM, Pinto BDL, Almeida LOB, Slaets JFW, Koberle R (2010) Recording from two neurons: second-order stimulus reconstruction from spike trains and population coding. \emph{Neural Comput} {\bf 22:} 2537--2557.

\bibitem{bialek+zee_1990}
Bialek W, Zee A (1990) Coding and computation with neural spike trains. \emph{J Stat Phys} {\bf 59:} 103--115.

\bibitem{rad+paninski_2011}
Rad KR, Paninski L (2011) Information rates and optimal decoding in large neural populations. \emph{Adv Neural Proc Syst} {\bf 24:} 846--854.

\bibitem{boerlin+deneve_2011}
Boerlin M, Deneve S (2011) Spike-based population coding and working memory. \emph{PLOS Comput Biol} {\bf 7:} e1001080.

\bibitem{boerlin+al_2013}
Boerlin M, Machens CK, Deneve S (2013) Predictive coding of dynamical variables in balanced spiking networks. \emph{PLOS Comput Biol} {\bf 9:} e1003258.

\bibitem{deneve+chalk_16}
Deneve S, Chalk M (2016) Efficiency turns the table on neural encoding, decoding and noise. \emph{Curr Opin Neurobiol} {\bf 37:} 141--148.

\bibitem{averbeck+al_2006} 
Averbeck BB, Latham PE, Pouget A (2006) Neural correlations, population coding and computation. \emph{Nat Rev Neurosci} {\bf 7:} 358--366.

\bibitem{schneidman+al_2006}
Schneidman E, Berry MJ 2nd, Segev R, Bialek W (2006) Weak pairwise correlations imply strongly correlated network states in a neural population. \emph{Nature} {\bf 440:} 1007-12.

\bibitem{ecker+al_2010}
Ecker AS et al. (2010) Decorrelated neuronal firing in cortical microcircuits. \emph{Science} {\bf 327:} 584--587.

\bibitem{granot+al_2013}
Granot-Atedgi E, Tka\v{c}ik G, Segev R, Schneidman E (2013) Stimulus-dependent maximum entropy models of neural population codes. \emph{PLOS Comput Biol} {\bf 9:} e1002922.

\bibitem{hyvarinen_2009}
Hyv\"arinen A, Hurri J, Hoyer PO (2009) Natural Image Statistics -- A probabilistic approach to early computational vision. Springer, London.

\bibitem{nelken_99}
Nelken I, Rotman Y, Yosef OB (1999) Response of auditory-cortex neurons to structural features of natural sounds. \emph{Nature} {\bf 397:} 154--156.


\bibitem{Vinje+Gallant_2000}
Vinje WE, Gallant JL (2000) Sparse coding and decorrelation in primary visual cortex during natural vision. \emph{Science} {\bf 287}: 1273--1276

\bibitem{Froudarakis+al_2014}
Froudarakis E et al. (2014) Population code in mouse V1 facilitates readout of natural scene through increased sparseness. \emph{Nat Neurosci} {\bf 17}: 851-860

\bibitem{Baudot+al_2013}
Baudot P et al. (2013) Animation of natural scene by virtual eye-movements evokes high precision and low noise in V1 neurons. \emph{Front Neural Circuits} {\bf 7}: 206

\bibitem{Hughes_1979}
Hughes, A. (1979) A schematic eye for the rat. \emph{Vision Res} 19(5):569-88.

\bibitem{Kim+al_2007}
Kim S-J, Koh K, Lustig M, Boyd S, and Gorinevsky D (2007)
An Interior-Point Method for Large-Scale l1-Regularized Least Squares.
\emph{IEEE J Sel Topics in Sign Proc}, 1(4):606-617

\bibitem{Lampert_2009}
Lampert CH (2009) Kernel Methods in Computer Vision. Foundations and trends in computer graphics and vision 4(3)

\bibitem{Bishop_2006}
Bishop CM (2006) Pattern Recognition and Machine Learning. Springer

\bibitem{McFarland_2013}
McFarland JM, Cui Y, Butts DA (2013) Inferring nonlinear neuronal computation based on physiologically plausible inputs. \emph{PLOS Comput Biol} {\bf 9:} e1003142.

\end{thebibliography}
\end{document}